\documentclass[12pt,preprint]{aastex}
\usepackage{graphicx}

\begin{document}

\title{Fast and Furious: Shock Heated Gas as the Origin of Spatially
  Resolved Hard X-ray Emission in the Central 5 kpc of the Galaxy
  Merger NGC~6240}

\author{Junfeng Wang\altaffilmark{1,2}, Emanuele
  Nardini\altaffilmark{1,3}, Giuseppina Fabbiano\altaffilmark{1},
  Margarita Karovska\altaffilmark{1}, Martin Elvis\altaffilmark{1}, Silvia Pellegrini\altaffilmark{4}, 
  Claire Max\altaffilmark{5}, Guido Risaliti\altaffilmark{1,6}, Vivian U\altaffilmark{7,8}, Andreas Zezas\altaffilmark{1,9}}

 \altaffiltext{1}{Harvard-Smithsonian Center for Astrophysics, 60 Garden St, Cambridge, MA 02138}
 \altaffiltext{2}{Center for Interdisciplinary Exploration and Research in Astrophysics (CIERA) and Department of Physics and Astronomy, Northwestern University, 2131 Tech Dr, Evanston, IL 60208}
 \altaffiltext{3}{{\it Current address:} Astrophysics Group, School of Physical and Geographical Sciences, Keele University,
Keele, Staffordshire ST5 5BG, UK}
 \altaffiltext{4}{Dipartimento di Astronomia, Universit\'{a} di Bologna, Via Ranzani 1, 40127 Bologna, Italy}
\altaffiltext{5}{Center for Adaptive Optics, University of California, 1156 High Street, Santa Cruz, CA 95064}
\altaffiltext{6}{INAF-Arcetri Observatory, Largo E, Fermi 5, I-50125 Firenze, Italy}

 \altaffiltext{7}{Institute for Astronomy, University of Hawaii, 2680 Woodlawn Drive, Honolulu, HI 96822}
 \altaffiltext{8}{{\it Current address:} Department of Physics and Astronomy, University of California, Riverside, CA 92521}
 \altaffiltext{9}{Physics Department, University of Crete, P.O. Box 2208, GR-710 03, Heraklion, Crete, Greece}

\email{jfwang@northwestern.edu}

\begin{abstract}

We have obtained a deep, sub-arcsecond resolution X-ray image of the
nuclear region of the luminous galaxy merger NGC 6240 with {\em
  Chandra}, which resolves the X-ray emission from the pair of active
nuclei and the diffuse hot gas in great detail.  We detect extended
hard X-ray emission from $kT\sim 6$ keV ($\sim$70 million K) hot gas
over a spatial scale of 5 kpc, indicating the presence of fast shocks
with velocity of $\sim$2200 km s$^{-1}$.  For the first time we obtain
the spatial distribution of this highly ionized gas emitting
\ion{Fe}{25}, which shows a remarkable correspondence to the large
scale morphology of H$_2$(1-0) S(1) line emission and H$\alpha$
filaments.  Propagation of fast shocks originated in the starburst
driven wind into the ambient dense gas can account for this
morphological correspondence.  With an observed $L_{0.5-8{\rm
    keV}}=5.3\times 10^{41}$ erg s$^{-1}$, the diffuse hard X-ray
emission is $\sim$100 times more luminous than that observed in the
classic starburst galaxy M82.  Assuming a filling factor of $1\%$ for
the 70 MK temperature gas, we estimate its total mass ($M_{\rm
  hot}=1.8\times 10^8 M_{\odot}$) and thermal energy ($E_{\rm
  th}=6.5\times 10^{57}$ ergs).  The total iron mass in the highly
ionized plasma is $M_{\rm Fe}=4.6\times 10^5 M_{\odot}$.  Both the
energetics and the iron mass in the hot gas are consistent with the
expected injection from the supernovae explosion during the starburst
that is commensurate with its high star formation rate.  No evidence
for fluorescent Fe I emission is found in the CO filament connecting
the two nuclei.
\end{abstract}


\section{Introduction}

In the local universe, NGC 6240
\citep[$z$=$0.02448\pm0.00003$;][]{Downes93} is a unique galaxy, in
the throes of a violent merging event and on its way to becoming an
elliptical galaxy \citep{Tacconi99,Tecza00,Bush08,Engel10,
  Medling11}. It is experiencing intense star formation (e.g., $61\pm
  30 M_{\odot}$ yr$^{-1}$ in Yun \& Carilli 2002; $25 M_{\odot}$ yr$^{-1}$ in Engel et al.\ 2010). With $L_{FIR}\sim 10^{11.8} L_{\odot}$ just below $10^{12}L_{\odot}$, the luminosity threshold for the ultraluminous infrared galaxies (ULIRGs, Sanders \& Mirabel 1996; Genzel et al.\ 1998), NGC 6240 is expected to become an ULIRG when the galaxies coalesce and a final starburst is triggered \citep{Engel10}.

The {\em Chandra} X-ray images \citep{Lira02,Komossa03} of the central
region of NGC 6240 revealed two gravitationally interacting active
supermassive black holes (SMBHs) separated by 1.5\arcsec\/ ($\sim$0.7
kpc).  These sources are characterized by the highly absorbed hard
X-ray spectra typical of Compton thick active galactic nuclei
\citep[AGNs;][]{Iwasawa98,Matt00}. They each have an observed luminosity of $L_{2-8{\rm
    keV}}\sim 10^{42}$ erg s$^{-1}$, and show prominent neutral Fe
K$\alpha$ lines at 6.4 keV present in the spectra of both nuclei
\citep{Komossa03}.

Limited by the number of counts, marginal evidence of an H-like Fe line
in the spectral fit and extended hard X-ray emission in the morphology
was also suggested by \citet{Komossa03}.  This line emission could be
associated with a spatially resolved, multi-temperature hot gas
outflow, powered by the starburst activity that dominates the 0.5-3
keV X-ray emission \citep[$L_{0.1-2{\rm keV}}\sim 10^{43}$ erg
  s$^{-1}$;][]{Schulz98,Komossa03,Ptak03,Huo04,Grimes05}.  {\em
  XMM}-Newton observations \citep{Boller03,Netzer05} resolved the Fe K
line complex into three narrow lines, the neutral Fe K$\alpha$,
\ion{Fe}{25}, and a blend of \ion{Fe}{26} with Fe K$\beta$. Using a
three-zone model, \citet{Netzer05} further identified that the
emission lines of higher ionization ions (e.g., \ion{Fe}{25} line)
originated from the inner 2.1 kpc region.  These {\em XMM}-Newton
data, however, do not have the resolution to resolve spatially the
line-emitting regions.

In this paper, we focus on the spatially resolved 5.5-8 keV X-ray
emission of the innermost central $\sim$5 kpc ($\sim 10\arcsec$)
region in NGC 6240 based on new, and archival, {\em Chandra} imaging.
This choice of hard energy range allows us to investigate the origin
of the high temperature gas and the highly ionized iron emission. A
detailed study of the extended soft X-ray halo has been presented in a
companion paper \citep{Nardini13}.  A third paper (Wang et al. 2013,
in preparation) will discuss the extended soft X-ray emission in the
$r \lesssim 15\arcsec$ region.  Taking advantage of the order of
magnitude deeper imaging, along with subpixel resolution, we are able
to firmly establish the presence and the spatial distribution of the
hard X-ray emission. Throughout this work we adopt a luminosity
distance of $D_L = 107$ Mpc to NGC 6240 (1\arcsec\/ = 492 pc), based
on the concordance cosmological parameters ($H_0$ = 70.5 km s$^{-1}$
Mpc$^{-1}$, $\Omega_m$=0.27, $\Omega_{\Lambda}$=0.73; Komatsu et
al. 2011).

\section{Observations and Data Reduction}

NGC 6240 was observed on May 31, 2011 with the {\em Chandra} Advanced
CCD Imaging Spectrometer \citep[ACIS;][]{Garmire03} for 146 ks (ObsID
12713, PI: Fabbiano).  The binary nucleus of NGC 6240 was placed near
the nominal aimpoint of the backside-illuminated, low energy sensitive
S3 chip.  The ACIS data were
reprocessed\footnote{\url{http://cxc.harvard.edu/ciao/threads/createL2/}}
with the {\em Chandra} Interactive Analysis of Observations software
package (CIAO; version 4.4) using the latest {\em Chandra} Calibration
Database (CALDB) version 4.5.1.  The net exposure time was 144 ks
after screening for brief periods of elevated background.  The
archival ACIS S3 observation \citep[ObsID 1590, previously presented
  in][]{Komossa03} shown in this work was taken on February 22, 2002
with a total exposure of 37 ks.  We reprocessed the ACIS data to
generate a new level 2 event file that has the latest calibration and
the subpixel resolution algorithm \citep[``EDSER'',][]{Li03} applied.
We checked for X-ray counterparts of the optical and near infrared
point sources in the field of view using the Naval Observatory
Mergered Astrometric Dataset catalog (NOMAD; Zacharias et al. 2004),
and compared the positions of all the bright point-like X-ray sources
across the observations. We verified the accuracy of the absolute
astrometry to within $<$1 native pixel (0.492 arcsec), thus no further
relative offsets were applied. The event files were all reprojected to
the coordinate frame of ObsID 12713, and the four images were merged
using CIAO tools {\tt reproject\_events} and {\tt dmmerge}.  Because
the effective area of the {\em Chandra} High Energy Transmission
Grating (HETG) zeroth-order image is about 60\% of ACIS-S at 6.4 keV,
we also make use of two archival deep HETG observations (PI:
Canizares; ObsID 6908, 159 ks, observed on May 16, 2006; ObsID 6909,
143 ks, observed on May 11, 2006), providing an equivalent of 180 ks
ACIS-S imaging in the Fe K band.  Therefore, the combined depth of our
imaging data reaches $\sim$363 ks, which improves by an order of
magnitude over the first ACIS image of NGC 6240 (Komossa et al. 2003)
in the 5.5-8 keV range. The first-order Chandra HEG spectrum is quite
faint, which was previously presented in Shu et al. (2011) but only
focused on measurement of the intrinsic width of the neutral iron
K$\alpha$ emission line. We obtained the HEG spectra from the TGCat
database (http://tgcat.mit.edu/) and combined the first orders of the
grating data (the positive and negative arms). The extraction aperture
was $r=4$ arcsec centered on the brighter northern nucleus. Given the
complex diffuse emission, using a larger extraction window will yield
unreliable Redistribution Matrix File (RMF) because of blurring of
spatial-spectral information along the dispersion direction. For
reference, the spectral resolution of HEG is $\Delta E=0.041$ keV at
6.4 keV ($\Delta
\lambda=0.012\AA$\footnote{\url{http://cxc.harvard.edu/proposer/POG/html/HETG.html}},
FWHM) and the systematic uncertainty in the HEG wavelength
calibration\footnote{\url{http://space.mit.edu/CXC/calib/hetgcal.html}}
is $\sim$0.01 keV ($\sim$430 km s$^{-1}$) at 6.4 keV. Table~\ref{tab0}
summarizes the {\em Chandra} ACIS and HETG observations used in our
work.

\section{Imaging Spectroscopic Analysis of the NGC 6240 Central Region}\label{sec:sec3}

\begin{figure}
\epsscale{1.0}
\plotone{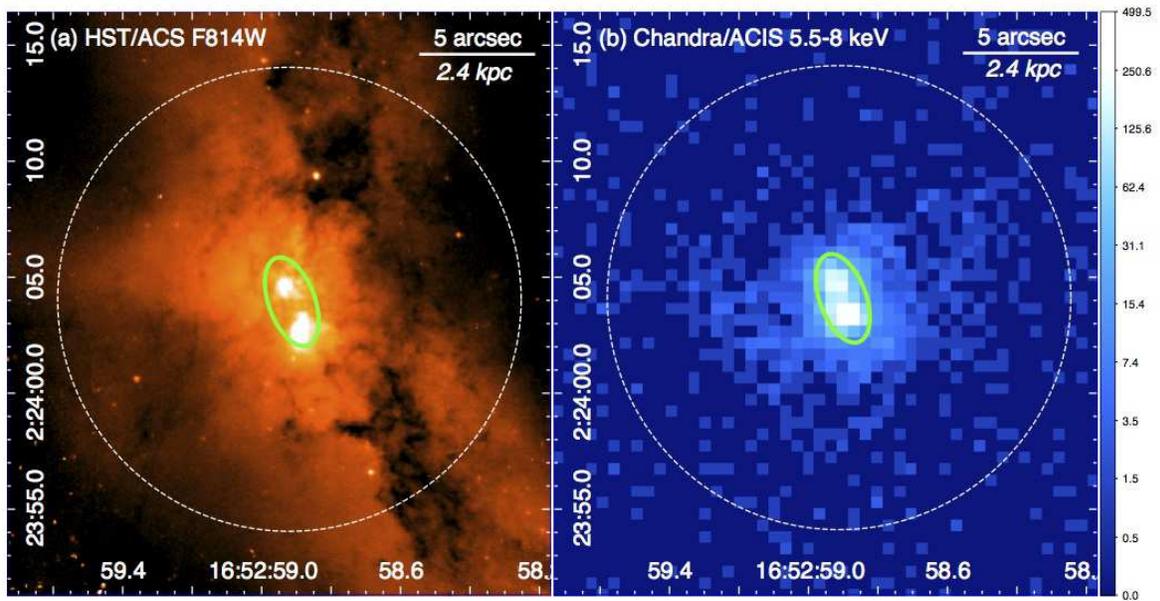}
\epsscale{1.0}
\caption{(a) HST/ACS F814W image of the central $25\arcsec \times
  25\arcsec$ (12 kpc across) region of NGC 6240, showing the disturbed
  optical disk. (b) The Chandra ACIS image (5.5-8 keV) of the same
  region extracted from the merged data.  The white circles
  ($r=10\arcsec$) mark the spatial extent of the hard X-ray
  emission.  The green ellipses ($4\arcsec\times 2\arcsec$) enclose
  the two nuclei.
\label{fig1}}
\end{figure}

Figure~\ref{fig1} shows the {\em Chandra} ACIS image of the inner $25\arcsec \times 25\arcsec$ ($\sim$12 kpc across) region of NGC 6240 in the 5.5-8 keV band extracted from the merged
data, together with an archival {\em HST} ACS/WFC F814W image (Cycle
14 program 10592, PI: A. Evans), highlighting the highly disturbed
optical morphology of NGC 6240. The two nuclei are resolved as the
brightest X-ray point sources in the center, with peak emission
separated by 1.5\arcsec\/ (710 pc), matching the nuclear compact radio
source positions (e.g., N1 and N2 in the 8.4 GHz map of Colbert et
al.\ 1994; N1 and S in the 1.4 and 5 GHz maps of Beswick et
al.\ 2001; North and South black hole positions in Max et al. 2007).

\begin{figure}
\plotone{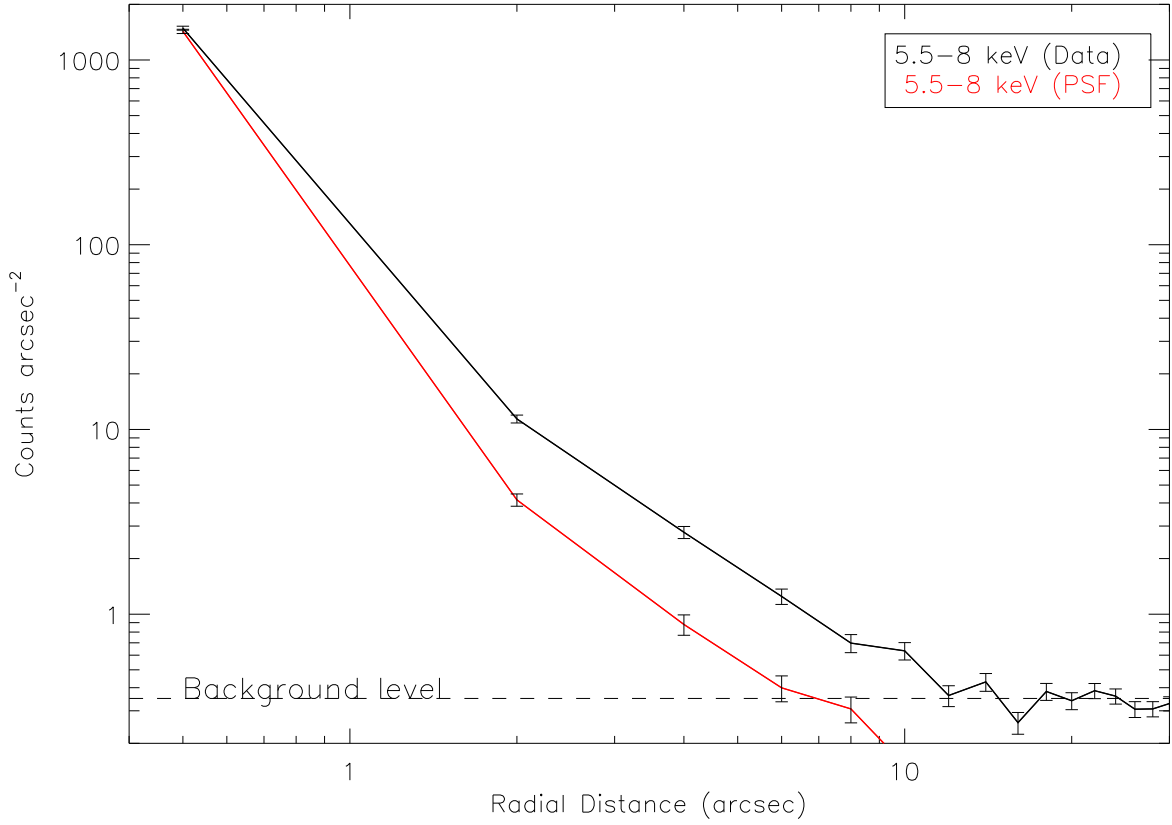}
\caption{Comparison between the observed surface brightness profile (black; full merged data)
  and that expected from the Chandra PSF for the 5.5-8 keV emission (red).
  Note the presence of significant hard X-ray emission reaching $r\sim
  10\arcsec$ radial distance.
\label{fig1b}}
\end{figure}

Diffuse emission is clearly present (Figure~\ref{fig1}b), extending to $r=10\arcsec$
($\sim$5 kpc). Hereafter we refer this region as the central region of NGC
6240. The X-ray morphology resembles a ``heart'' with loop-like
features in the south, the north-east, and the north-west directions (Figure~\ref{fig1}b).
A comparison between the observed surface brightness profile and that
of a point source further confirms the presence of significant hard
X-ray emission reaching $r\sim 10\arcsec$ radial
distance. Figure~\ref{fig1b} shows the excess of hard X-rays over the
scattered emission in the wings of the point spread function
(PSF). The radial profile flattens past 10\arcsec\/, where background
emission dominates.  The {\em Chandra} PSF shown in Figure~\ref{fig1b}
was created simulating a point source at the same position on the
detector as the center of NGC 6240 in the single observation of the
longest exposure (ObsID 12713) using the {\em Chandra} Ray Tracer
(ChaRT\footnote{\url{http://cxc.harvard.edu/chart/}}) and the
MARX\footnote{Version 4.4.0; \url{http://space.mit.edu/cxc/marx/}}
simulator. This PSF was then rescaled to match the combined 5.5-8 keV
counts from the two nuclei. Since the nuclei are separated by
1.5\arcsec\/, and we are interested in demonstrating the presence of
extended emission several arcseconds from the nuclei, this is an
adequate procedure. For the investigation of extended emission close
to the nuclei (e.g., between the nuclei), we have performed a more
complex 2-D image fitting with two PSF components, described in the
next section.

\subsection{Spectral Analysis: Neutral, He-like and H-like Fe K lines}

Previous work by Boller et al. (2003) identified ionized Fe K emission
lines from the presence of a collisionally ionized plasma with a
temperature $kT=5.5\pm 1.5$ keV, producing the \ion{Fe}{25} (at 6.7
keV) and \ion{Fe}{26} (at 6.93 keV) K$\alpha$ recombination
lines. Boller et al. also inserted in their fit the neutral or low
ionization Fe K$\alpha$ line (expected at 6.4 keV) and K$\beta$ line
(at 7.058 keV), each modelled with a Gaussian line. Such fluorescent
lines of large (up to few keV) equivalent widths (EWs) arise from
large column density gas which strongly attenuates directly viewed
X-rays from the AGNs.

\begin{figure}
\centerline{ \includegraphics[scale=0.6,angle=-90]{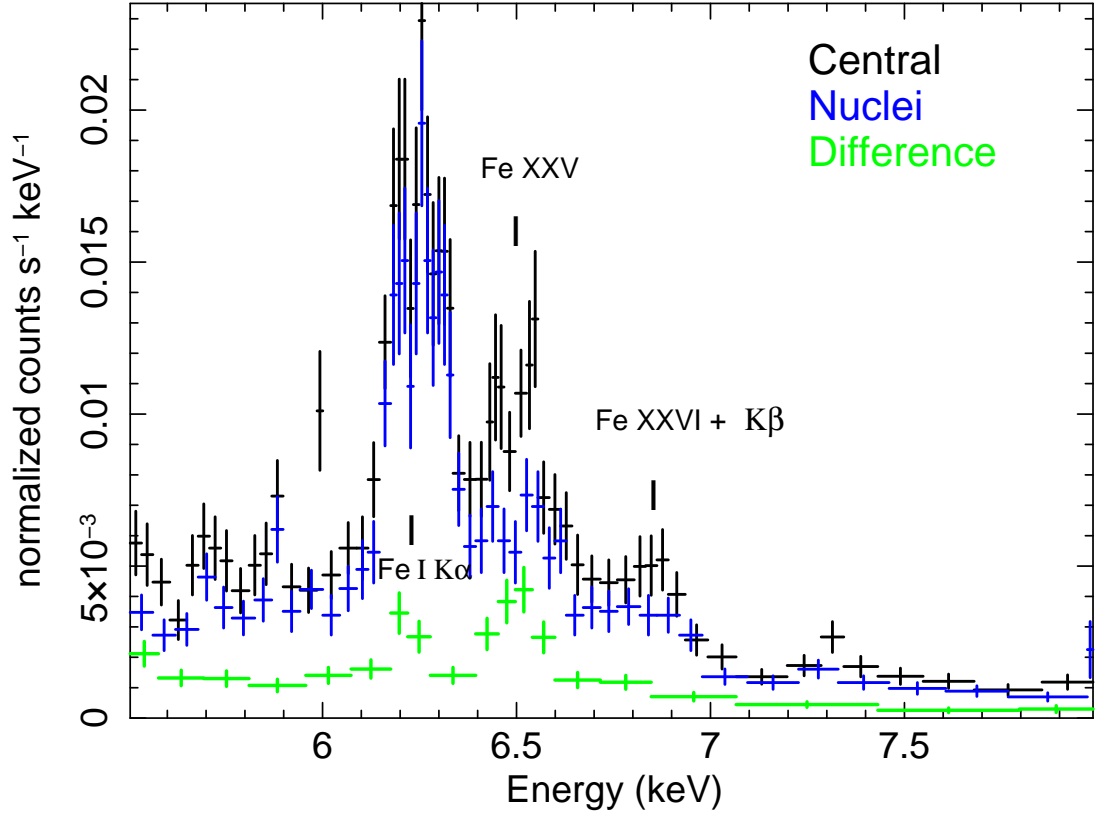}}
\caption{Comparison between the X-ray spectra for the dual AGN
  region (shown in blue; extracted from the ellipse in
  Figure~\ref{fig1}) and the 10 arcsec radius region (shown in black;
  extracted from the circle in Figure~\ref{fig1}) including both the
  diffuse and AGN emission.  The X-ray spectrum for the central
  region excluding the nuclei is also shown (green) to illustrate the spectral difference and its flux level. The energy scale is observed (not rest frame).
\label{fig2a}}
\end{figure}

To shed light on the spectral properties of the spatially resolved central region, we
extracted the spectra of both nuclei and all the emission in the
10\arcsec-radius circle.  Figure~\ref{fig2a}a compares the ACIS
spectra of the central region of NGC 6240 (circle in
Figure~\ref{fig1}, all emission within $r$=10\arcsec, $\sim$5 kpc from
the nuclei) and the innermost region containing the two nuclei
(ellipse in Figure~\ref{fig1}, full lengths of the axes $4\arcsec\times 2\arcsec$).

The complex Fe K line structure detected in {\em XMM}-Newton spectra
\citep{Boller03,Netzer05} is clearly present in the {\em Chandra} ACIS
spectra (Figure~\ref{fig2a}) and contributes most counts, in contrast
to the low continuum in the same region.  Approximately 80\% of this
continuum comes from the innermost region enclosing the nuclei.  There
are notable differences between the two spectra, which are further
illustrated in Figure~\ref{fig2a} showing the residual spectrum
excluding the nuclei.  Following \citet{Boller03}, we added four
narrow Gaussian emission lines (representing neutral Fe K$\alpha$,
\ion{Fe}{25}, \ion{Fe}{26}, and Fe K$\beta$) to a highly absorbed
power law continuum to fit the spatially resolved spectra (the
northern nucleus, the southern nucleus, and the extended emission
within $r=10\arcsec$, i.e., nuclei+diffuse).  This phenomenological
approach is adopted to identify the spatial distribution of the line
emission, and a more physically meaningful model will be used in the
later section.  To prevent oversubtraction of diffuse emission from
the hot gas in the nuclear region, we did not exclude the innermost
region when evaluating the hot gas content in the $r\leq 10\arcsec$
emission; instead, we included nuclear components to represent the
contribution from the AGNs. 

Figure~\ref{fig2a} also shows a hint for two peaks in the \ion{Fe}{25}
line, which resembles double peaked emission line profiles seen in the
optical spectra of AGN with bi-conical outflows \citep[e.g. Mrk
  78,][]{Whittle04}. However, it is statistically insignificant: the
probability that the model with one more gaussian component improves
over the current fit is only 80\% based on a likelihood ratio test
(10000 simulations) performed using XSPEC. In fact, the energy
resolution of the ACIS S3 detector is 0.2
keV\footnote{\url{http://cxc.harvard.edu/proposer/POG/html/chap6.html}}
(FWHM) at 6.4 keV whereas the apparent double peaks are separated by
0.05 keV, therefore we suggest that it is not a real resolved feature.
The \ion{Fe}{26} and K$\beta$ lines are blended together at the
spectral resolution of the ACIS CCD, therefore we fixed the K$\beta$
line center energy at 7.058 keV (rest frame) and scaled the line
intensity according to the theoretical ratio \citep[1/8.8,
  e.g.,][]{Palmeri03} relative to the K$\alpha$ line.  The best-fit
centroid energies, EWs, and line fluxes, and associated 1-$\sigma$
errors are summarized in Table~\ref{tab1}.

Although the first-order {\em Chandra} HEG spectrum is faint, we
attempted to spectroscopically resolve the various Fe features. The
grating spectrum was rebinned for a minimum S/N ratio of 3 and was
fitted with the similar model consisting of an absorbed power law
continuum plus gaussian lines using XSPEC. Again we caution that the
best-fit value of $\Gamma$ for the simplistic power law continuum
model is not physically meaningful but simply parameterization. We
recovered a $FWHM=2860\pm 550$ km s$^{-1}$ for the FeK$\alpha$ 6.4 keV
line, consistent with the $FWHM\sim 2810$ km s$^{-1}$ reported in Shu
et al. (2011).  Interestingly, the FeXXV line at 6.7 keV is marginally
resolved at $2.4\sigma$ significance with the measured
$FWHM=5550_{-1390}^{+1670}$ km s$^{-1}$, implying a large velocity
range in the hot gas kinematics. Furthermore, the HEG spectrum does
not resolve this line into two peaks, although it shows the line is
broad ($FWHM\sim 5550$ km s$^{-1}$). More counts would have allowed
finer binning to attempt resolving the broad line into two or more
kinematic components.

The line fluxes further confirm that the \ion{Fe}{25} emission is
spatially extended while the neutral Fe K$\alpha$ emission is
concentrated at the nuclei.  The neutral or weakly ionized Fe
K$\alpha$ line emission at 6.4 keV (centered at 6.25 keV in the
observer frame) is also mostly associated with the nuclei, with
$\sim$15\% of the total Fe I K$\alpha$ emission in the extended region
(Figure~\ref{fig2a}).  At the 6.4 keV energy, the PSF scattering of
nuclear emission to the larger radii cannot be neglected, which is
$\sim$10\%.  Taking this into account, the spatially extended Fe I
K$\alpha$ emission is at 5\% of the total.  The \ion{Fe}{26} emission
is fully accounted for by the innermost nuclear emission, indicating
that the mostly ionized H-like iron is confined in this region.  In
contrast, $\sim$30\% of \ion{Fe}{25} line emission appears to
originate in the extended region outside of the nuclear region (taking
into account of the PSF contribution). In the following we will
investigate the morphology of this emission.  Given that the XMM
spectra \citep{Netzer05} come from a larger region, the similarity of
the fluxes with our {\em Chandra} measurements excludes significant Fe
emission beyond $r=10\arcsec$.

\subsection{Narrow-band Hard X-ray Images}

\begin{figure}
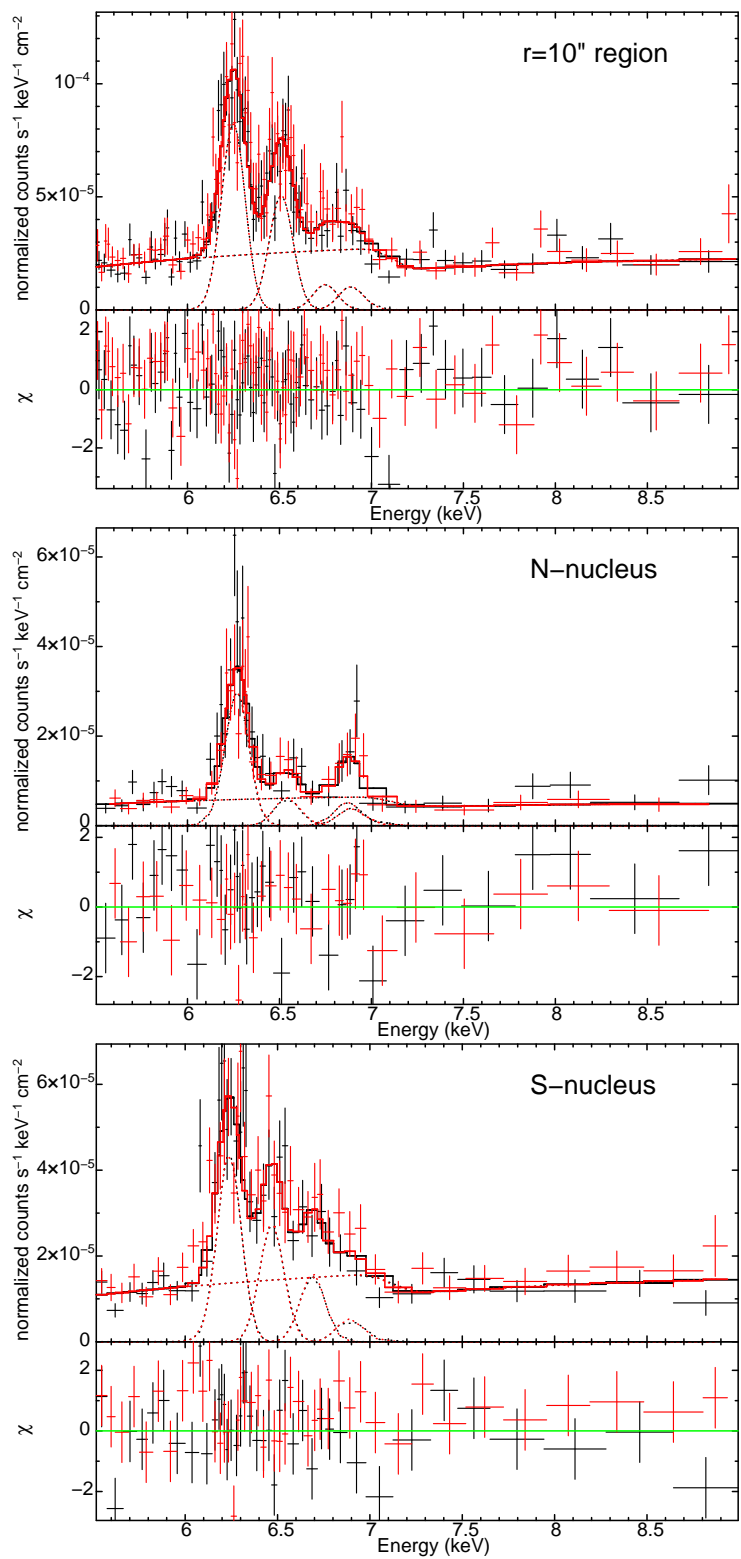

\centerline{\includegraphics[scale=0.4,angle=-90]{f4a.eps}}
\centerline{\includegraphics[scale=0.4,angle=-90]{f4b.eps}} 
\centerline{\includegraphics[scale=0.4,angle=-90]{f4c.eps}}
\caption{Top to bottom: ACIS spectra and fits for the extended central
  emission, the northern nucleus, and the southern nucleus.  Both ACIS
  data (black) and HETG (red) zeroth order data were used.  The model
  components are also shown as dashed lines (see Table~\ref{tab1}).
\label{fig2}}
\end{figure}

The analysis above shows that the AGNs alone in NGC 6240 cannot
account for the observed Fe K complex, except for the Fe K$\alpha$ and
K$\beta$ emission (the latter fainter and blended with \ion{Fe}{26}),
which is due to fluorescence of neutral or low ionization material
near the AGNs.  Guided by the spectra (Figures~\ref{fig2a}
and~\ref{fig2}), we extracted {\em Chandra} narrow-band line images to
investigate the morphology of the emission line gas.

\begin{figure}
\centerline{\includegraphics[scale=1.07,angle=90]{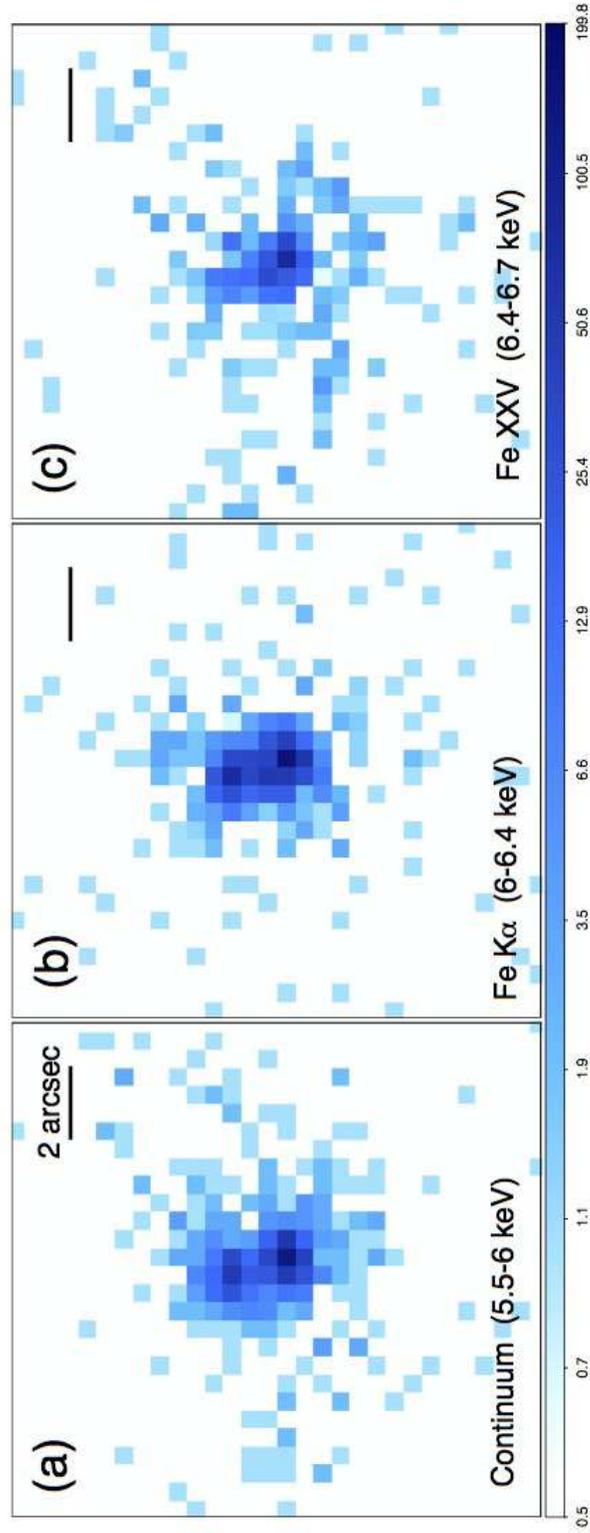}} 
\caption{Narrow band ACIS images: (a) the 5.5-6 keV continuum band;
  (b) Fe K$\alpha$ line (6-6.4 keV); and
  (c) \ion{Fe}{25} line (6.4-6.7 keV). The latter two images were continuum-subtracted (see text).
\label{fig3}}
\end{figure}

Figure~\ref{fig3} shows images in the 5.5-6 keV continuum band that is
free from any significant line emission, the Fe K$\alpha$ line (6-6.4
keV) and the \ion{Fe}{25} line (6.4-6.7 keV). The blended \ion{Fe}{26}
and Fe I K$\beta$ emission was examined but not shown here, since it
does not show any evidence for emission beyond the nuclear region.
Although the line emission dominates in these narrow bands, the
contribution of the continuum has been removed from the images
regardless, assuming a single power law with a photon index
($\Gamma=0.8$) from the spectral fit of the continuum in the central
region.  The contribution from the continuum in the emission line
bands is calculated using the photon index and the counts in 5.5-6.0
keV, taking into account the difference in the effective areas.  Given
that the starburst likely dominates the extended emission
\citep[e.g.,][]{Netzer05}, we further confirmed that the following
results do not change adopting a thermal bremsstrahlung spectrum
(kT$\sim$6 keV for the hot gas) to approximate the flat continuum.

The continuum-subtracted line emission images clearly show that the
6.4 keV Fe I K$\alpha$ line peaks at the dual nuclei, with weak
emission extending to $r=3\arcsec$ from the nuclei and contributing
$\sim$10\% of the total $\sim$900 Fe K$\alpha$ counts. The
\ion{Fe}{25} emission is also centrally peaked at the nuclear region
with a dominating southern nucleus, but it appears more extended to
the southwest and to the east; the $r=10\arcsec$ region outside of the
nuclei contains $\sim$40\% of the $\sim$500 total \ion{Fe}{25} counts.
These values are in good agreement with the results from spectral
analysis above.

\begin{figure}
\centerline{\includegraphics[scale=1.07,angle=90]{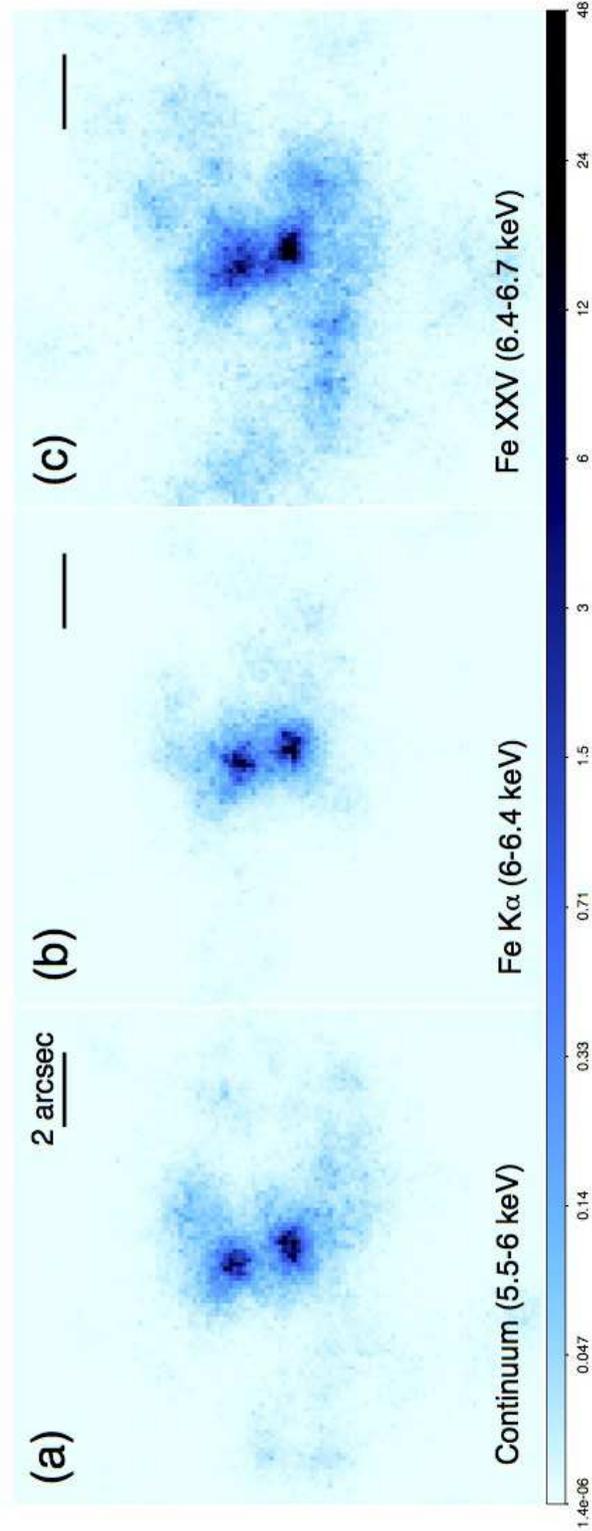}} 
\caption{Narrow band ACIS images, with EMC2 PSF-deconvolution applied:
  (a) the continuum band (5.5-6 keV); (b) line image for the neutral
  Fe K$\alpha$ line (6-6.4 keV); and (c) the \ion{Fe}{25} (6.4-6.7
  keV).
\label{fig4}}
\end{figure}

To extract more morphological information from the data, we
performed image restoration using the Expectation through Markov Chain
Monte Carlo (EMC2) algorithm \citep{Esch04,Karovska05}.  The
effectiveness of this method with {\em Chandra} images was
demonstrated with a number of astronomical imaging studies
\citep{Karovska05, Karovska07, Wang09a, Wang12}.  The EMC2
deconvolution is more effective for preserving diffuse features
compared to other PSF-deconvolution algorithms
\citep[e.g.,][]{Lucy74}. Only the single deep imaging observation
(ObsID 12713) and the associated PSF were used to minimize the PSF
variations across multiple observations.  The PSF-deconvolved images,
shown in Figure~\ref{fig4}, provide the highest resolution view of the
central region.  In particular, the faint diffuse \ion{Fe}{25} emission
is more obvious in Figure~\ref{fig4}c.

\subsection{Sub-kpc-scale Extended Emission between the Nuclei}

\begin{figure}
\epsscale{0.75}
\plotone{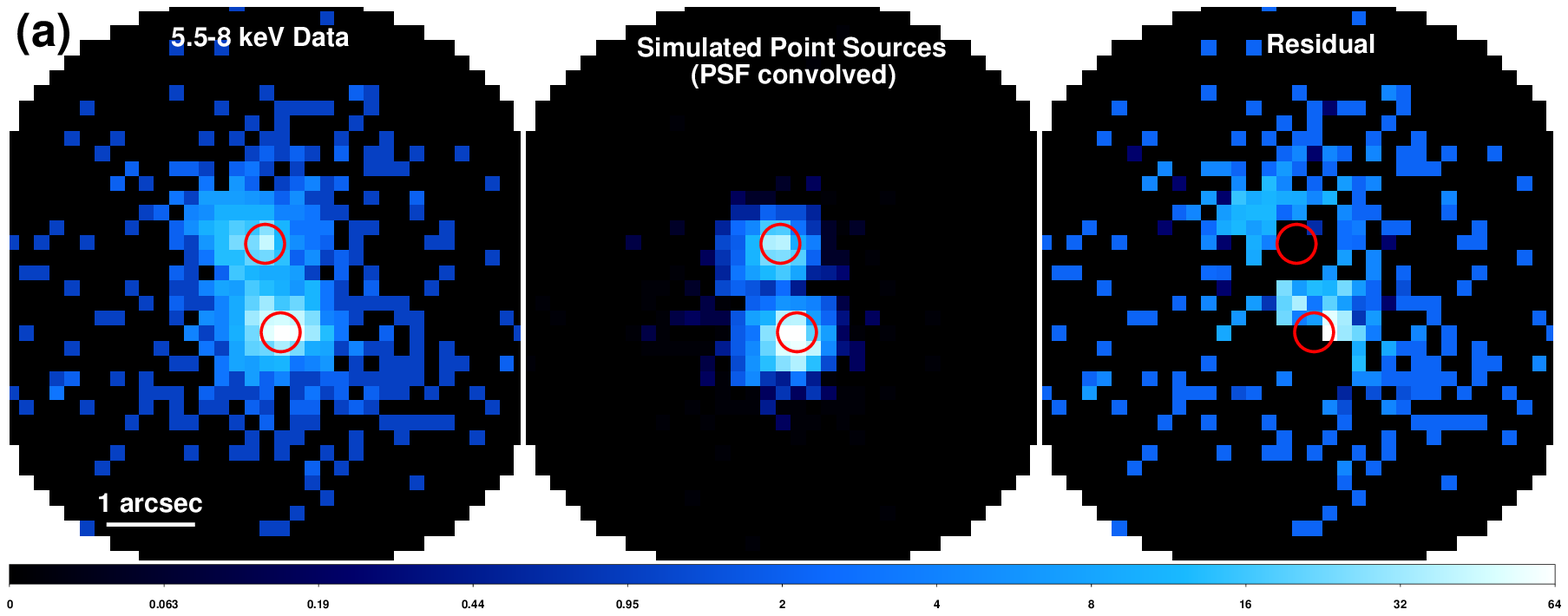}
\plotone{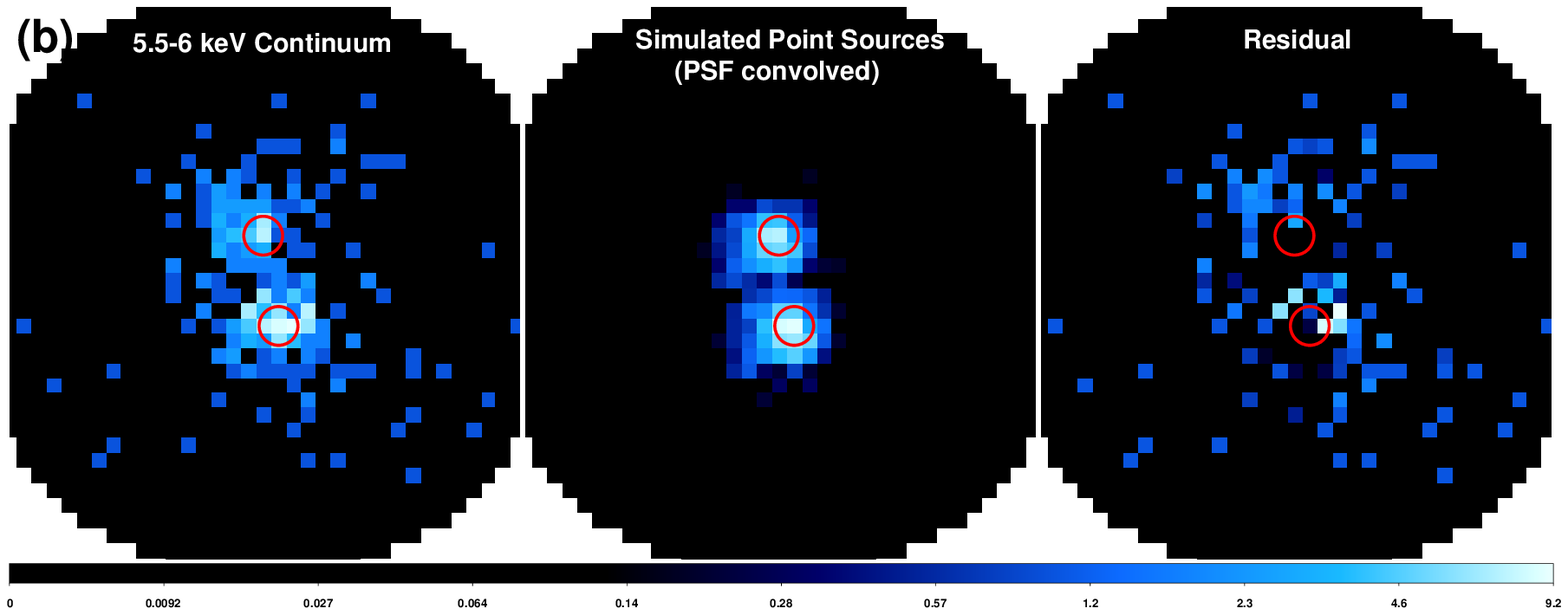}
\plotone{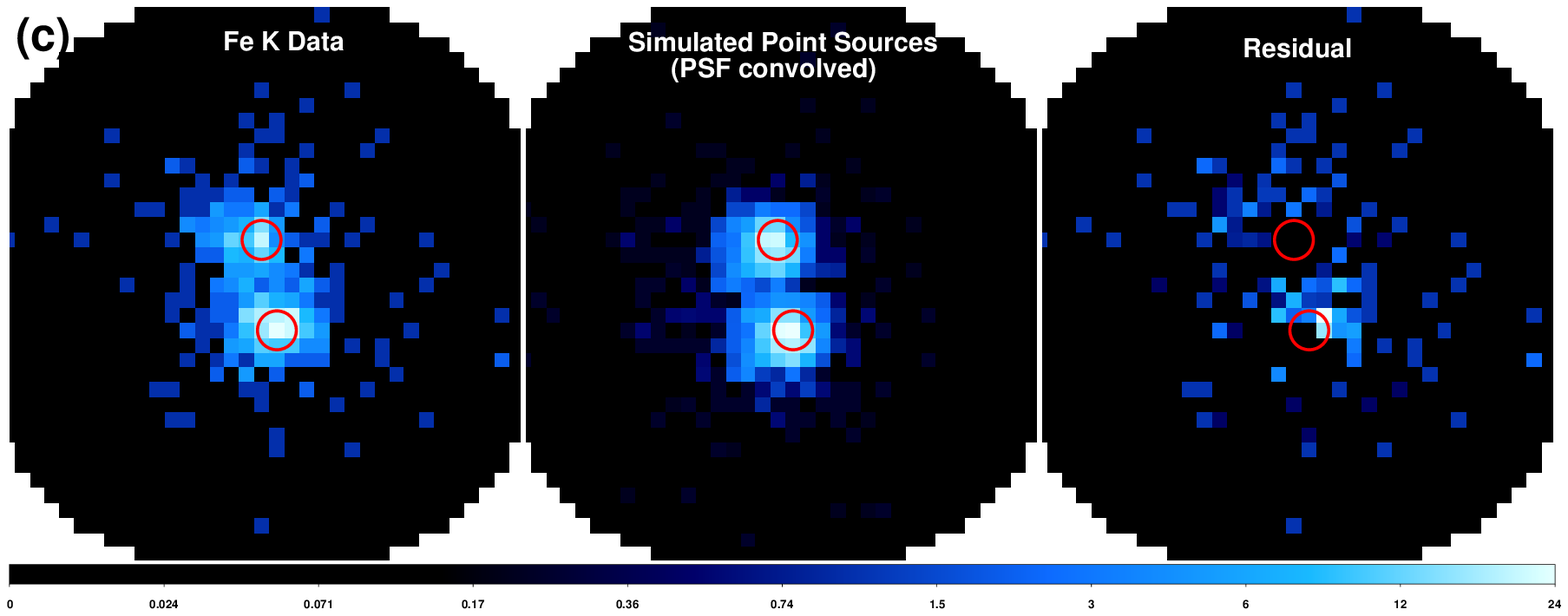}
\plotone{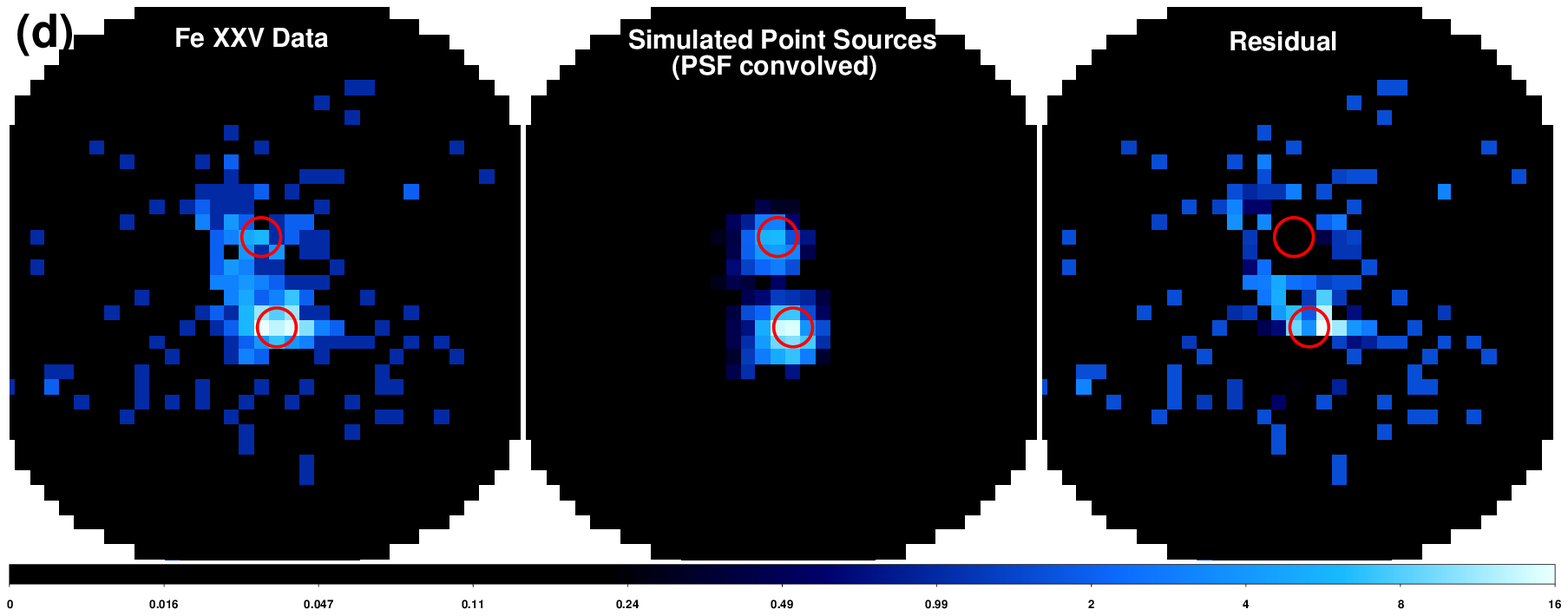}
\epsscale{1.0}
\caption{2-D image fitting of the innermost nuclear region in (a)
  5.5-8 keV (b) 5.5-6 keV (continuum) (c) 6.0-6.4 keV (Fe I) (d)
  6.4-6.7 keV (Fe XXV). For each band, the observed image, simulated
  pair of point sources representing the double nuclei, and the
  residual emission are shown. The two circles mark the compact radio
  nuclear sources \citep{Colbert94}.
\label{fig5}}
\end{figure}

Close examination of the images suggests the possibility of some
extended hard X-ray emission in the region between the two nuclei,
especially in the Fe XXV band (and perhaps in the Fe K$\alpha$ band;
see Figures~\ref{fig3} and~\ref{fig4}).  However, evaluating the
significance of this emission is very difficult because of the close
separation (1.5\arcsec\/) of the nuclei.  The PSF wings of the two
X-ray point sources overlap in this region and enhance the emission
here, which needs to be taken into account.  Simulating two point
sources convolved with the PSF, we performed 2-D image fitting to subtract the PSF\footnote{See detailed procedure available at \url{http://cxc.harvard.edu/sherpa4.4/threads/2dpsf/}.} on the
5.5-8 keV image, as well as individual narrow band images, using {\tt
  Sherpa}\footnote{Version 2 in CIAO 4.4;
  \url{http://cxc.harvard.edu/sherpa/}}.  The results are shown in
Figure~\ref{fig5}.  For the 5.5-8 keV image that has most counts,
besides the clear extension to the southwest and the northeast
(already noted above as the extended central region), the residual
image further indicates that there is extended emission in the region
between the two nuclei, immediate north of the southern nucleus, at
$\sim$10~$\sigma$ significance level when evaluated using a $1\arcsec
\times 1\arcsec$ box region.  For the narrow band images, the
statistics are poorer; nevertheless, this extended feature is seen in
all three residual images.  Its significance ranges from
$\sim$6~$\sigma$ in the \ion{Fe}{25} line emission to $\sim$3~$\sigma$
in the Fe K$\alpha$ line emission.

We further checked for possible contamination from the PSF
artifact\footnote{\url{http://cxc.harvard.edu/ciao4.4/caveats/psf\_artifact.html}}
found in ACIS data. Using the CIAO tool {\tt
  make\_psf\_asymmetry\_region}, this artifact should lie in position
angle $0-50^{\rm o}$ in our data.  Given that this PSF asymmetry
feature may contain only 5\% of the flux\footnote{See document written
  by Vinay Kashyap (CXC) available at
  \url{http://cxc.harvard.edu/cal/Hrc/PSF/acis\_psf\_2010oct.html}},
it would have in any case minimal impact on our results.

\subsection{Comparison with Multiwavelength Images}

In this section, we compare the hard X-ray morphology of the center of
NGC 6240 with a number of high spatial resolution images available in
other wavelength ranges, aiming to find clues for the physical origin
of the extended X-ray emitting gas.  The absolute astrometry of our
{\em Chandra} ACIS image is excellent, indicated by the agreement
between the coordinates of the X-ray and the radio nuclei
($\sim$0.2\arcsec\/ accuracy, see Max et al. 2007 for a further astrometry discussion).

\begin{figure}
\centerline{\includegraphics[scale=0.9,angle=0]{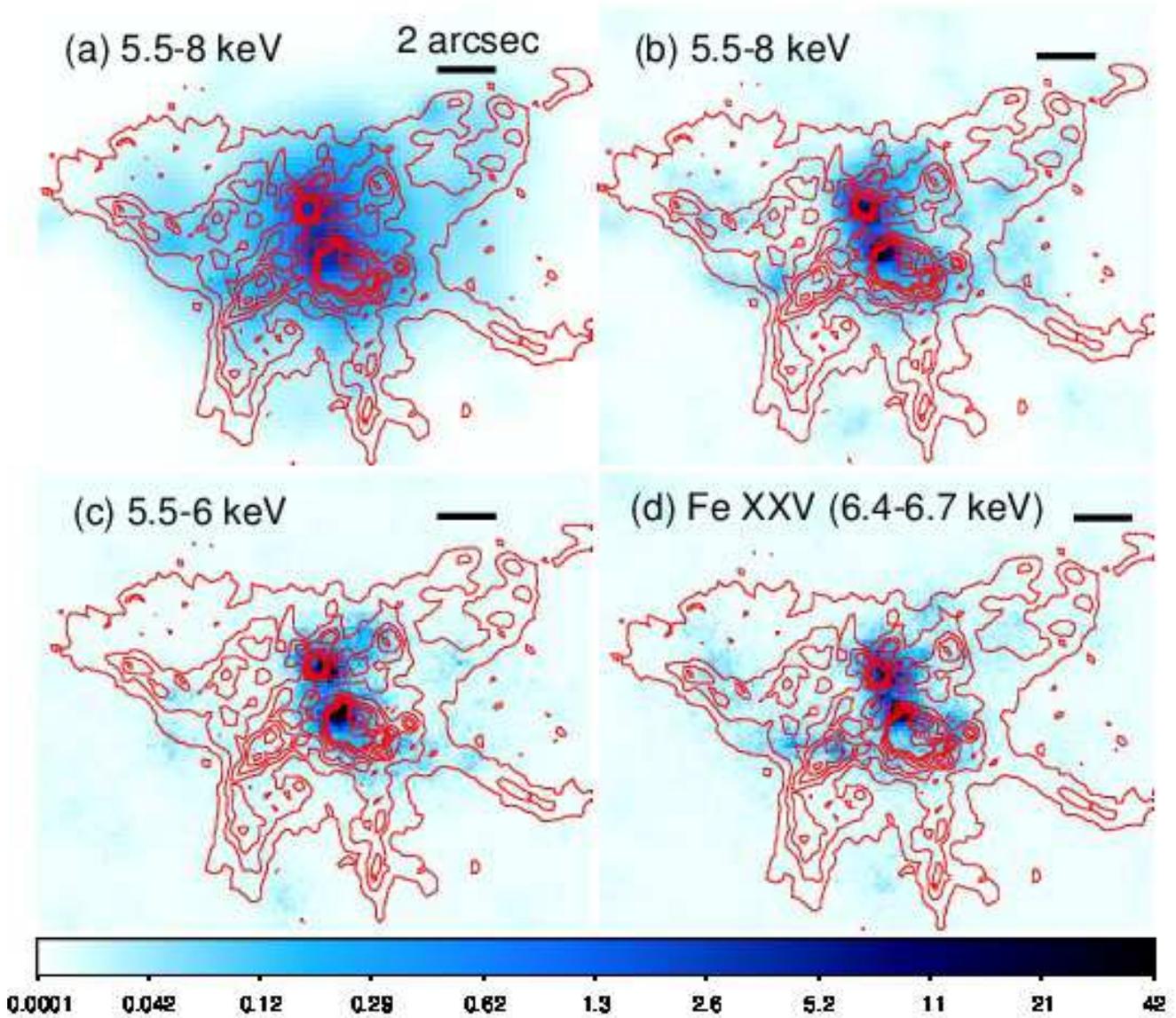}} 
\epsscale{1}
\caption{(a) The adaptively smoothed 5.5-8 keV ACIS image (in 1/2 pixel binning) overlaid with the HST/WFPC2 H$\alpha$ contours \citep{Lira02}. (b) The PSF-deconvolved 5.5-8 keV image overlaid with the same H$\alpha$ contours. (c) The PSF-deconvolved continuum band (5.5-6 keV) image (Figure~\ref{fig4}a) overlaid with the same H$\alpha$ contours. (d) The PSF-deconvolved \ion{Fe}{25} image (Figure~\ref{fig4}c) overlaid with the same H$\alpha$ contours. 
\label{fig6a}}
\end{figure}

We first compare the 5.5-8 keV emission with the optical H$\alpha$
emission morphology from {\em HST}/WFPC2 F673N image \citep{Lira02},
as shown in Figure~\ref{fig6a}a.  The 5.5-8 keV image was adaptively
smoothed with CIAO tool {\tt dmimgadapt} using a Gaussian kernel with
smoothing scales of 1 pixel to 20 pixels, to better visualize
structures in the diffuse emission.  As a separate approach, the same
5.5-8 keV image with the PSF deconvolution applied is also shown in
Figure~\ref{fig6a}b.  The optical peaks are not aligned with the X-ray
nuclei due to obscuration. \citet{Lira02} and \citet{Komossa03} have
previously shown that on 10\arcsec\/ (4.9 kpc) scale, the H$\alpha$
emission and the soft X-ray emission share a similar morpholgy
consisting of filaments and loops. The overall morphology of the soft
X-ray emission in our deep image (Wang et al. 2013) is
similar to that shown in Komossa et al.\ (2003) but with finer
structure details, closely following the optical H$\alpha$
emission. Such morphological correspondence is seen but less obvious
in the hard X-ray emission, except for the bright H$\alpha$ filament
to the east and the loop to the northwest of the nuclei.
Figure~\ref{fig6a}c and~\ref{fig6a}d further compare the H$\alpha$
morphology with images in the 5.5-6 keV continuum band and the
\ion{Fe}{25} band.  A close morphological correspondence between the
H$\alpha$ and the hard X-rays can be found at the base of an
incomplete loop to the west of the nuclei (Figure~\ref{fig6a}c and d),
and at the filament to the south and east of the southern nucleus
(Figure~\ref{fig6a}d).

Next we compare the 5.5-8 keV image with the [FeII] 1.644~$\mu$m line
emission \citep{vanderWerf93} in Figure~\ref{fig6}.  The [FeII]
emission in NGC 6240 could be produced by shocks in supernova remnants
(SNRs; see van der Werf et al.\ 1993 for detailed discussion on
possible excitation mechanisms), and is brightest at the locations of
the two nuclei, consistent with the presence of nuclear starbursts
\citep{Engel10}.  The two [FeII] emission peaks fall close to the
X-ray nuclei, with a slight offset (0.2\arcsec\/) to the north
relative to the X-ray northern nucleus.

We find a close correlation of the spatially extended \ion{Fe}{25}
emission with the near IR H$_2$(1-0) 2.12~$\mu$m emission obtained
with Keck II \citep{Max05}.  Previous work \citep{vanderWerf93,Max05}
noted that the H$_2$(1-0) emission in NGC 6240 does not follow the
stellar light but peaks between the two nuclei, which is thought to be
excited by shocks \citep{Moorwood88,vanderWerf93}.  To compare H$_2$
and our X-ray images, the astrometry of the H$_2$ image was first
registered using an IR-bright star (2MASS J16530119$+$0224138,
$K$=10.4 mag) in the field.  Figure~\ref{fig7} shows a remarkably good
correspondence between the observed (top panel), and especially the
deconvolved (lower panel) Fe XXV line emission and the H$_2$(1-0)
2.12~$\mu$m emission.

Finally we compare the Fe I K$\alpha$ emission with the cold molecular
gas distribution in the nuclear region, traced by the CO(3-2) emission
obtained with the Submillimeter Array (SMA) from U et al. (2011).
Figures~\ref{fig8}a and~\ref{fig8}b show that the CO emission
resembles a `bridge' or filament between the two nuclei, with the
bright peak offset to the north of the southern Fe I K$\alpha$ peak.
The evidence for fluorescent Fe I emission from this cold gas is
marginal at most, with Fe K$\alpha$ emission extending to the north of
the southern nucleus at 3~$\sigma$ significance level.

\subsection{Spectral Fit with Thermal Plasma Model}\label{sec:physical}

To extract the physical parameters of the highly ionized gas
responsible for the \ion{Fe}{25} and \ion{Fe}{26} emission, we fitted
the 5.5--9~keV spectra of the $r\leq 10\arcsec$ region.  The
contribution from the lower temperature gas (e.g., $kT=0.7$ keV and
$kT=1.4$ keV components; Boller et al. 2003) found in previous studies
is negligible in this energy range.  We adopted an absorbed single
temperature thermal plasma ($APEC$; Smith et al. 2001) model for the
collisionally ionized gas producing the highly ionized \ion{Fe}{25}
and \ion{Fe}{26} emission lines, plus zero-width Gaussian lines
centered at 6.400 keV and 7.058 keV (rest frame) as the neutral Fe
K$\alpha$ and K$\beta$ lines, respectively.  A photoionized origin for
the \ion{Fe}{25} was not considered as this was previously ruled out
by the detailed photoionization calculation given by \citet{Netzer05}.
The continuum contribution from the two nuclei is accounted for by
including an absorbed power law component, representing the reflected
nuclear emission. The gaussian line components were also subject to
this absorption column.  The absorption column ($N_H=4\times 10^{23}$
cm$^{-2}$), the photon index ($\Gamma=0.9$) and the normalization
($A=1.5\times 10^{-4}$) are fixed to be the best fit values combining
the two nuclei.  The abundance $Z$ is fixed at solar values
\citep{grsa98}.

The spectrum with the best-fit model is plotted in Figure \ref{fig9},
which is well represented with a $\chi^2/d.o.f=300/246$.  The
temperature of the hot plasma is $6.15\pm 0.33$ keV with an absorption
column of $N_H=5.5\pm 1.7 \times 10^{23}$ cm$^{-2}$.  The observed
0.5--8 keV luminosity is $5.3\pm 1.0\times$10$^{41}$~$\mathrm{erg~
  s^{-1}}$ and $L_{0.5-8{\rm keV}}=5.5\pm
1.4\times$10$^{42}$~$\mathrm{erg~ s^{-1}}$ after correction for
absorption.  The normalization is $2.5\pm0.5\times 10^{-3}$, in units
of [$10^{-14} / (4\pi (D_{\rm A} (1+{\rm z}))^2) \int n_{\rm e}~
  n_{\rm H}~ {\rm dV}$], where $D\rm _A$ is the angular size distance
to NGC 6240 (cm), $n\rm _e$ is the electron density (cm$^{-3}$), and
$\rm n_H$ the hydrogen density (cm$^{-3}$).  We obtain an emission
measure (E.M.$\equiv \int n_e n_H \eta dV$) of $3.1\times 10^{65}
\eta$ cm$^{-3}$, where $\eta$ is a volume filling factor.

It is unlikely that all the hot plasma emission is affected by a high
column, given the clumpy morphology and the spatial
extent. Fortunately, because we are working in the hard X-ray range,
the absorption corrected luminosity is less vulnerable to the
uncertainty in the absorption column. Previous studies (Boller et
al. 2003) and the broad band fit indicate an absorption column of
$N_H=4.1\times 10^{22}$ cm$^{-2}$ is reasonable outside of the nuclear
region. This implies that our absorption corrected luminosity could be
over-estimated by a factor of 2 (or 0.3 dex in $\log L_x$).  To
investigate the impact of the soft X-ray thermal components that are
clearly present (Komossa et al.\ 2003, Netzer et al.\ 2005), we
modeled the broad band spectrum of the extended emission (the same
$r\leq 10\arcsec$ region). Figure~\ref{fig9a} shows the 0.3--8 keV
band ACIS spectrum (grating zero-order spectra were not combined
because of low effective area in the soft X-ray band; see discussion
in Nardini et al. 2013). The best fit model components are overplotted
including two more thermal components with lower temperature
($kT_1=1.03$ keV, $kT_2=1.56$ keV), each subjected to an absorption
column ($N_{H,1}=3.3\times 10^{21}$ cm$^{-2}$, $N_{H,2}=5.3\times
10^{22}$ cm$^{-2}$). The high temperature component has a similar
$kT_3=6.0$ keV that is consistent with the previous fitted
value. Fixing their abundances at solar values gives statistically
unacceptable fit ($\chi_ {\nu} >3$) and allowing for variation relative
to solar abundance improves the fit to ($\chi_ {\nu}=1.32$).  Remaining
residuals are a few peaks higher than model predicted line emission at
line energies corresponding to MgXII (rest frame 1.47 keV), SiXIV
(2.01 keV) and SXVI (2.62 keV), indicating likely anomalous abundances
for these elements. Indeed Netzer et al. (2005) suggested these
elements have $\sim$2 times higher abundance. A word of caution is
that the soft X-ray emission over a large spatial scale suffers from
blending of thermal plasma of inhomogeneous absorption, temperature,
and abundance, therefore spatially resolved studies in small regions
is preferred. For our purpose of evaluating the contamination from
these components to the hard energy range, we only add in gaussian
lines to improve the fit ($\chi_\nu =1.18$) which are shown in
Figure~\ref{fig9a}. We evaluated that the declining hard tails of
the soft X-ray spectral components only contaminate the 5.5--8 keV
spectral range at 8\% level, which has a minimal impact on the derived
$N_H$ ($4.0\pm 0.3\times 10^{23}$ cm$^{-2}$ when the hard tails from
the softer X-ray components are accounted for).

Using the \ion{Fe}{25} line emission as a tracer for the extent of the
hottest thermal component, which is confined within the central $r=10
\arcsec$, we can obtain the volume of the thermal gas emitting hard
X-rays, $V=1.5\times 10^{67} \eta$ cm$^3$, assuming a spherical
geometry.  This implies $n\rm _e =0.15 \eta^{-\frac{1}{2}}$ cm$^{-3}$.
Following the calculations done in Richings et al. (2010), the 70
million K gas contains a total mass of $M_{hot}=1.8\times 10^9
\eta^{\frac{1}{2}} M_{\odot}$ and a thermal energy of
$E_{th}=6.5\times 10^{58} \eta^{\frac{1}{2}}$ ergs.  Since we assumed
solar abundance, this implies the total iron mass in the highly
ionized plasma $4.6\times 10^6 \eta^{\frac{1}{2}} M_{\odot}$.  Note
that we also explored fitting with abundance left free and the best
fit model gives $Z=0.5^{+0.3}_{-0.2} Z_{\odot}$ with a lower
$kT=4.9\pm 1.1$ keV, which has minor impact on the above estimated
values.  In addition, there is a degeneracy between the flat powerlaw
component (which is most effectively contrained by the very hard end
of the spectrum) and the hot thermal component (most effectively
constrained by the \ion{Fe}{25} line assuming fixed solar
abundance). We find that a 0.05 dex increase in the normalization for
the powerlaw component corresponds to a 0.09 dex decrease in the
thermal component.  When non-solar abundance is allowed, it leads to a
0.19 dex decrease in the normalization for the thermal component. This
has negligible impact to our discussion later on the energetics of the
hot gas.

The filling factor of the hot gas is poorly constrained
observationally.  Nevertheless, hydrodynamic simulations by Strickland
\& Stevens (2000) suggest that $\eta$ ranges between 0.1\% to 10\% for
X-ray bright galactic winds.  In the starburst region $\eta$ is
considered to be higher than in the large scale wind, close to 100\%
(Strickland \& Heckman 2007).  To facilitate our discussion, we assume
$\eta \sim 1\%$ and caution the large uncertainty on this value.  We
consider that this is reasonable average value for the large volume
spanned by the hot gas; the true filling factor is most probably close
to unity in the innermost X-ray brightest region and $\ll 1\%$ in the
outer sphere.  However, we also note that the dependence on $\eta$ is
relatively weak for the hot gas mass and thermal energy
($\eta^{\frac{1}{2}}$).

\section{Discussion: Origin of the Diffuse Hard X-rays}

We have unambiguously detected diffuse hard (5.5-8 keV) X-ray emission
in the central region of the merging galaxy NGC 6240.  The Fe K$\alpha$ and
K$\beta$ fluorescent lines in NGC 6240 have been firmly associated
with the photoionized gas of high column density and low ionization
close to the AGN (e.g., Komossa et al. 2003, Netzer et al. 2005).
Most of the observed \ion{Fe}{1} line flux is explained by the
emission from the NGC 6240 nuclei, and the spatial concentration at
the nuclear positions is consistent with this interpretation. Even with
the long exposure, we do not find strong evidence that the dense
molecular clouds in the nuclear region are producing fluorescent Fe
lines.  The excess of Fe I K$\alpha$ counts in the CO bridge between
the nuclei (Figure~\ref{fig8}) is marginal (3$\sigma$ significance)
when the scattered counts from both AGN are excluded.

What is the origin of the highly ionized gas emitting the \ion{Fe}{25}
line that extends over 5 kpc in NGC 6240?  Given the LIRG nature of
NGC 6240 characterised by a high SFR (25 M$_{\odot}$ yr$^{-1}$, Engel
et al.\ 2010), we first estimate the contribution to hard X-ray
emission by the ensemble of X-ray binary (XRB) systems in the
starburst.  Using the correlation between the galaxy-wide 2--10 keV
luminosity and the SFR provided by the {\em Chandra} survey of LIRGs
(Lehmer et al.\ 2010, Table 4), the expected $L^{gal}_{HX}$ from the
XRB population is $3.3\times 10^{40}$ erg s$^{-1}$ in the 2--10 keV
band, adopting a $SFR\sim 25$~M$_{\odot}$ yr$^{-1}$ (Engel et
al.\ 2010).  This is only 5\% of the observed 2--10 keV luminosity of
the extended hard X-ray emission, strongly indicating that the diffuse
emission is dominated by hot gas.  Such hard X-ray emission is most
likely associated with the thermal gas from merged supernova (SN)
ejecta and stellar winds present during a starburst.  The He-like
\ion{Fe}{25} line emission is also observed in other well known
starburst galaxies like NGC 253 and M82
\citep[e.g.][]{Pietsch01,Mitsuishi11}, the ULIRG Arp 220
\citep{Iwa05}, and the integrated spectrum of a number of LIRG/ULIRG
systems \citep{Iwa09}, suggesting the existence of similar high
temperature plasma.  It is worth noting that the observed diffuse hard
X-ray emission in NGC 6240 here is almost 100 times more luminous than
that observed in the classic superwind system M82 \citep[$L_{2-8{\rm
      keV}}=4\times 10^{39}$ erg s$^{-1}$;][]{Strickland07}.

Using optical imaging and spectroscopy to identify outflow in the warm
ionized medium, Heckman et al. (1990) have shown NGC 6240 to host a
superwind.  Superwinds are believed to be driven by the thermal and
ram pressure of an initially very hot ($T\sim 10^8$ K), high-pressure
($P/k \sim 10^7$ K cm$^{-3}$) and low-density wind. According to the
starburst driven superwind model, a hot gas bubble of internally
shocked wind material with a temperature of several keV forms in the
region of intense starformation (Chevalier \& Clegg 1985, Suchkov et
al.\ 1994, Strickland \& Heckman 2007); this hot gas eventually flows
outward as a high-speed (few 1000 km s$^{-1}$) wind.  Here we detected
and spatially constrained for the first time thermal X-ray emission
from such hot ($T\sim 70$ MK) gas in NGC 6240, likely the thermalized
SN ejecta within the vicinity of the starburst region.

To further investigate whether the thermal energy content of the hot
gas could be powered by thermalization of SNe shocks, we need to
estimate the kinetic energy input from the SNe during the starburst.
The resulting heating by SNe shocks should at least be comparable to
$E_{th}$, considering that there are lower temperature phases and the
gas has also bulk kinetic energy.  The most often quoted SN rate for
NGC 6240 is $\sim 2$ yr$^{-1}$
\citep[e.g.,][]{vanderWerf93,Beswick01,Pollack07}, which will be used
next, although we note there is a small range of estimates.  Engel et
al.\ (2010) calculated a lower SN rate of 0.3 yr$^{-1}$ assuming 20
Myr of continuous star formation at 25 M$_{\odot}$ yr$^{-1}$, based on
the measured K-band luminosity.  Using the correlation between [FeII]
1.257~$\mu$m and the SN rate recently found by \citet{Rosenberg12}
from a sample of nearby starburst galaxies (Equation 2 and Figure 9
therein), we find a higher but comparable SN rate ($\sim$3 yr$^{-1}$).
Assuming a fraction ($\alpha=0.1$, Chevalier \& Clegg 1985, Thornton
et al. 1998) of the kinetic energy input ($10^{51}$ ergs per SN) is
converted into thermal energy of the hot gas, the total energy
deposited during the past $\Delta t_{SB}\sim 20$ Myr of most recent
starburst \citep{Tecza00} is $E=4\times 10^{57}$ ergs, which is
comparable to $E_{th}=6.5\times 10^{57}$ ergs.  Adopting a 10\%
filling factor for the hot gas results in a larger $E_{th}$ and a
larger deficiency ($E < E_{th}$ for $\alpha=0.1$), but it does not
impair our conclusion, considering that the thermalization efficiency
can be higher than the adopted value of 0.1 in an environment where
many SNe are exploding \citep{Strickland09}, which will yield a larger
$E$ that is still consistent with $E_{th}$.  These estimates
demonstrate that the observed diffuse thermal gas traced by the highly
ionized iron line emission in NGC 6240 is consistent with being heated
by SNe shocks in the starburst.  The cooling time of this thermal gas
is $\tau_{cool}=E_{th}/Lx \sim 40$ Myr, slightly larger than the
starburst duration, which implies that such X-ray bright phase will be
short lived once the starburst ceases.

Following \citet{Mitsuishi11}, another useful check is the comparison
between the ejected iron mass from SNe and the observed iron mass.
Adopting $M = 8.4\times 10^{-2} M_{\odot}$ as the ejected iron mass
per Type II SN \citep{Iwamoto99} and the average SN rate of $\sim 2$
yr$^{-1}$, the expected iron mass for the 20 Myr starburst is
$M_{Fe,SB}=3.4\times 10^{6} M_{\odot}$, which can easily account for
the observed $M_{Fe,X-ray}=4.6\times 10^{5} M_{\odot}$ considering not
all the iron ejecta are currently X-ray emitting.  Therefore this
comparison also supports the above scenario that starburst is the
origin for the 70 MK gas.

\begin{figure}
\epsscale{0.9}
\plotone{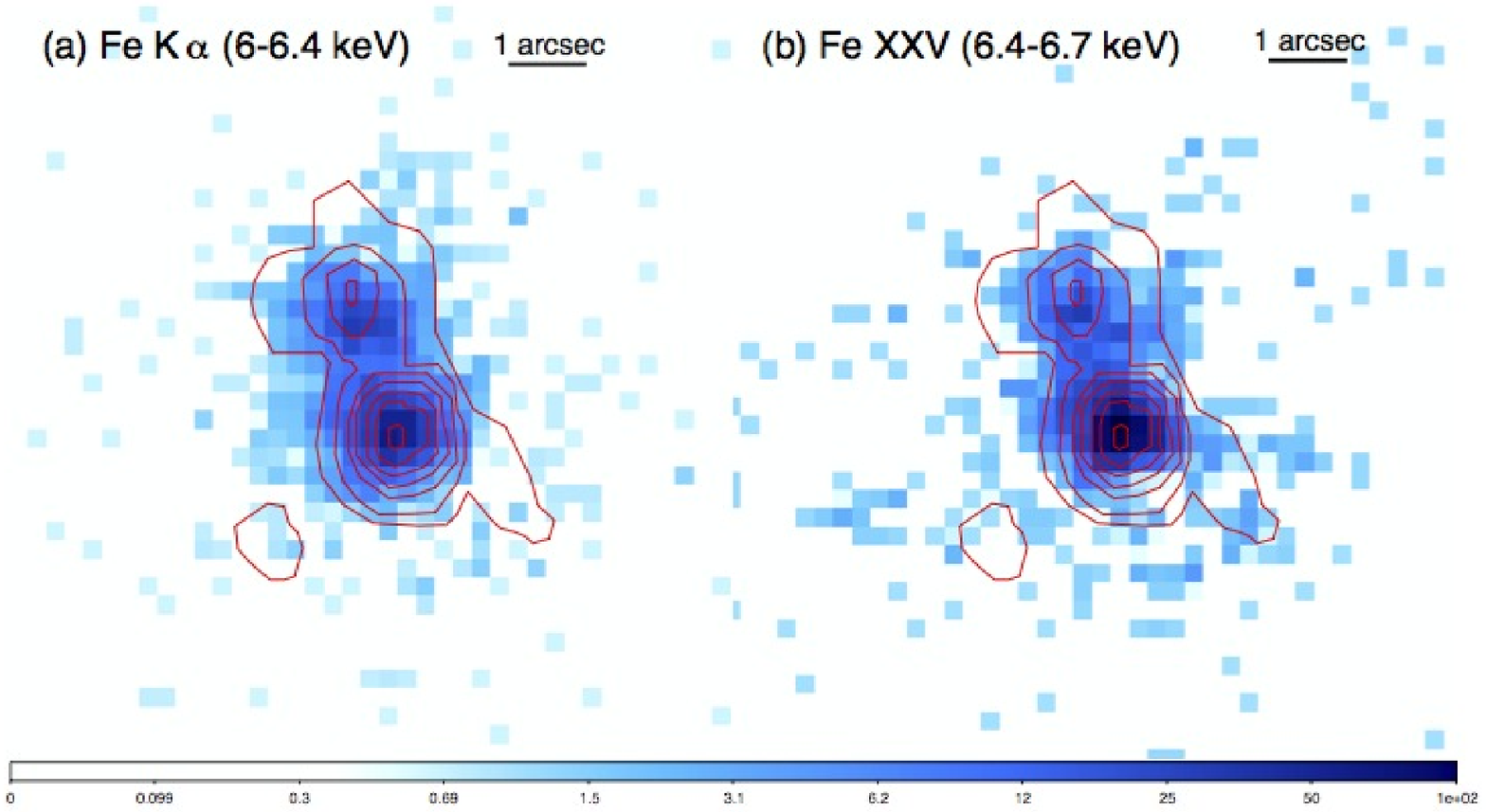}
\plotone{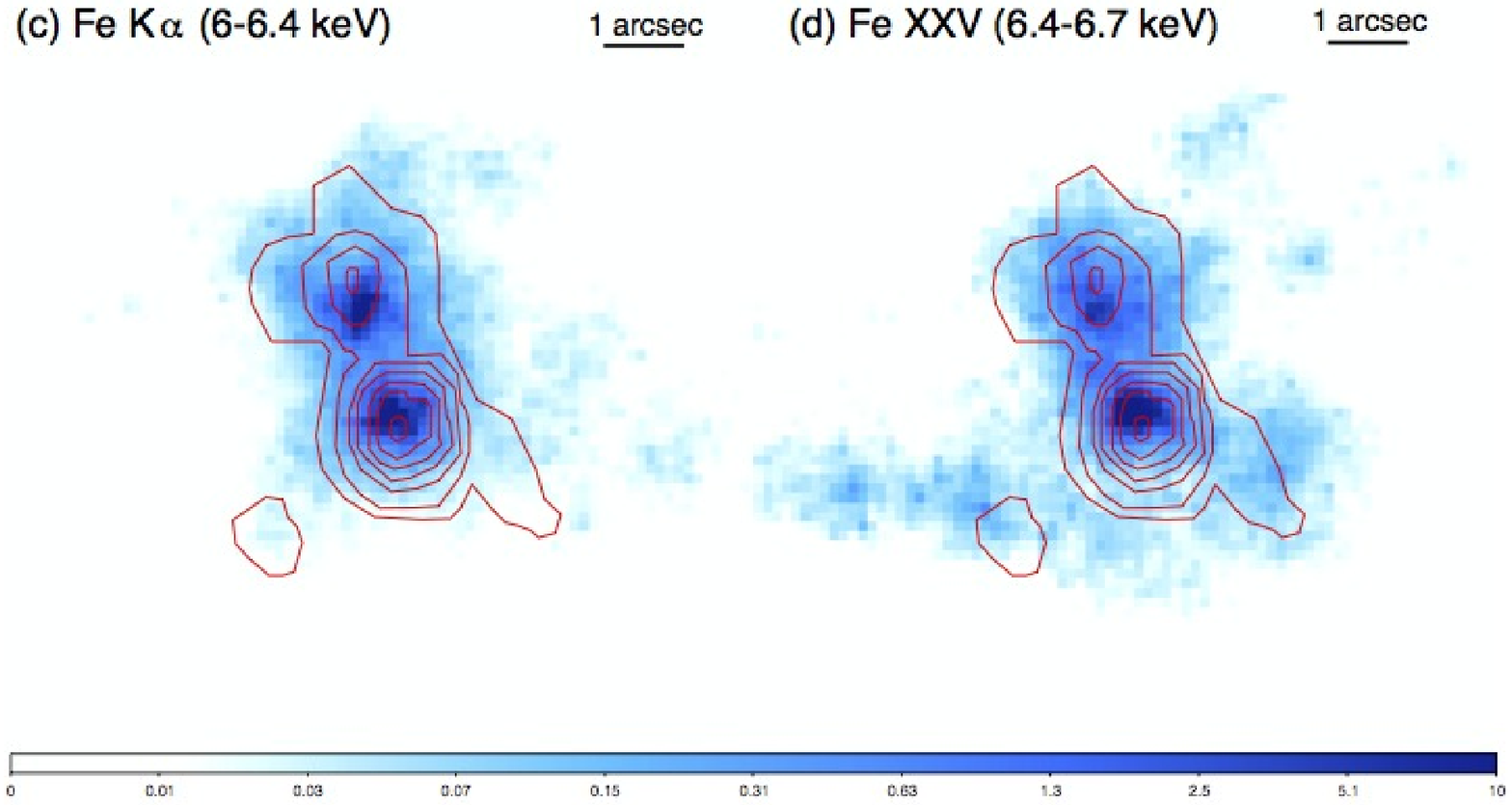}
\epsscale{1}
\caption{The near-IR [FeII] 1.64~$\mu$m contours (van der Werf et
  al. 1993) overlaid on: (a) raw (6-6.4 keV) Fe I K$\alpha$ emission;
  (b) raw (6.4-6.7 keV) \ion{Fe}{25} line emission; (c) same as (a)
  but with EMC2 deconvolution; (d) same as (b) but with EMC2
  deconvolution.
\label{fig6}}
\end{figure}

\begin{figure}
\epsscale{0.6}
\plotone{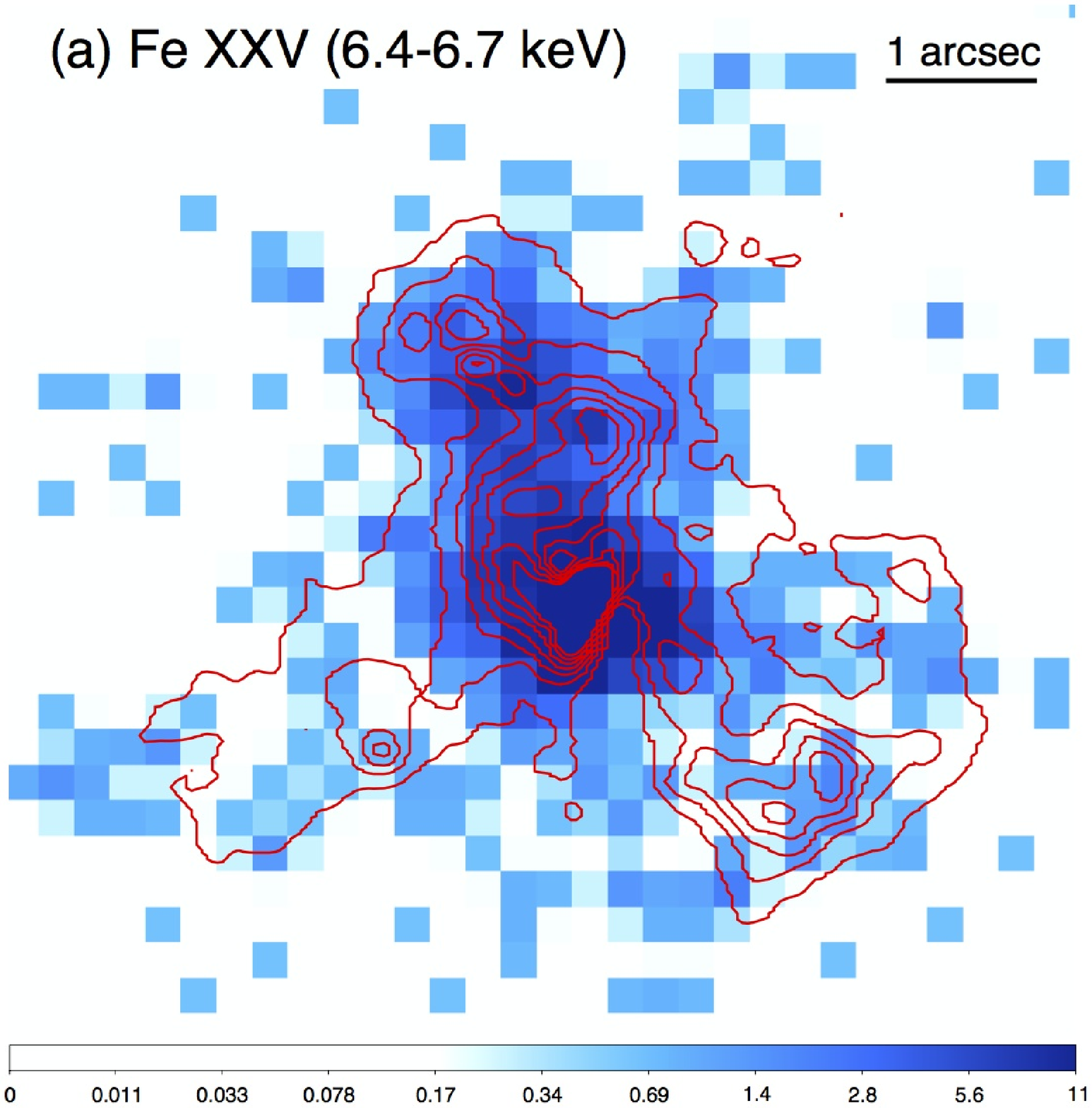}
\plotone{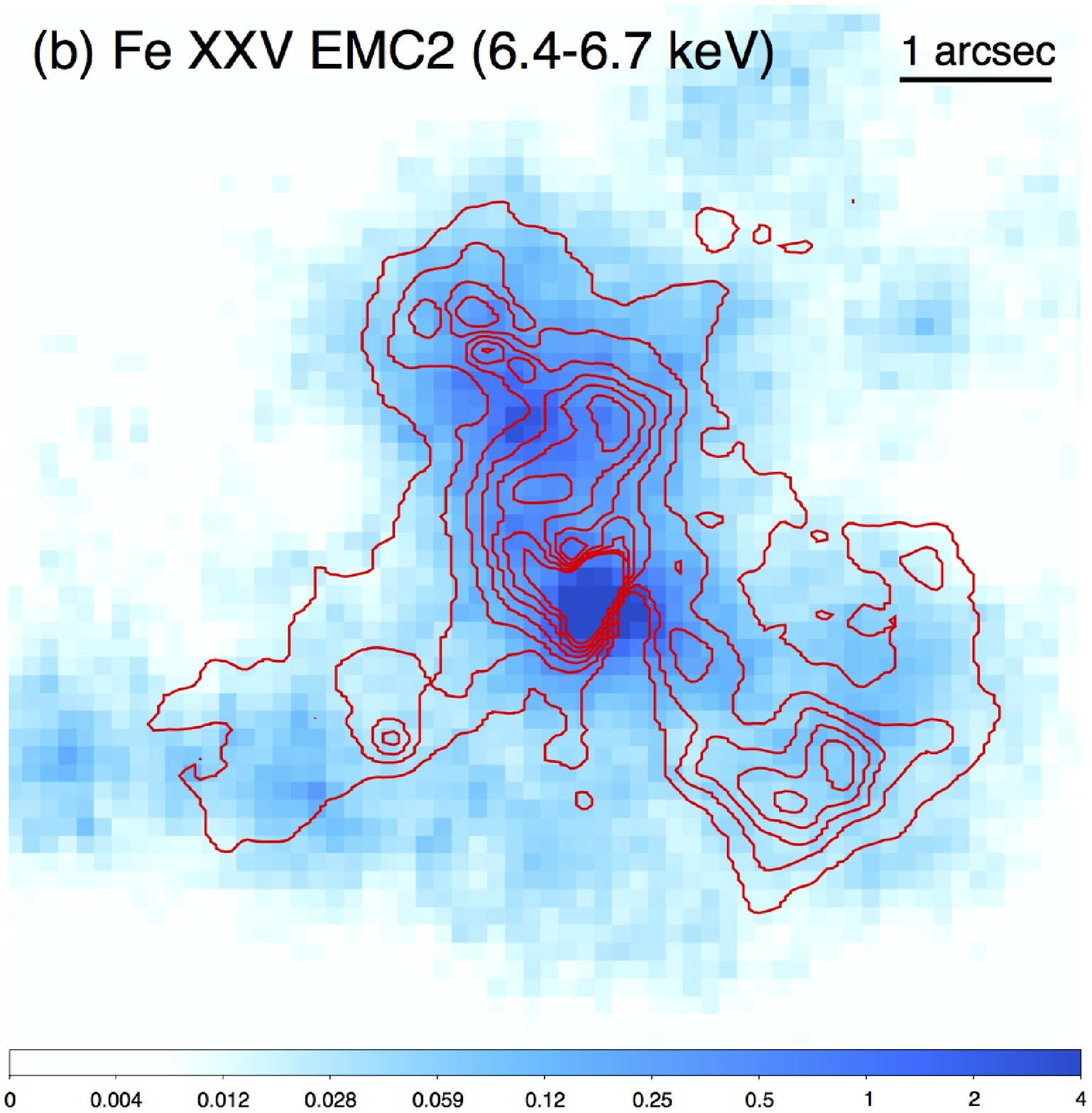}
\epsscale{1}
\caption{(a) The FeXXV line emission image overlaid with the contours
  of H$_2$(1-0) 2.12~$\mu$m emission (Max 2005); (b) PSF-deconvolved
  Fe XXV line emission image with the same H$_2$ contours.
\label{fig7}}
\end{figure}

\begin{figure}
\epsscale{0.6}
\plotone{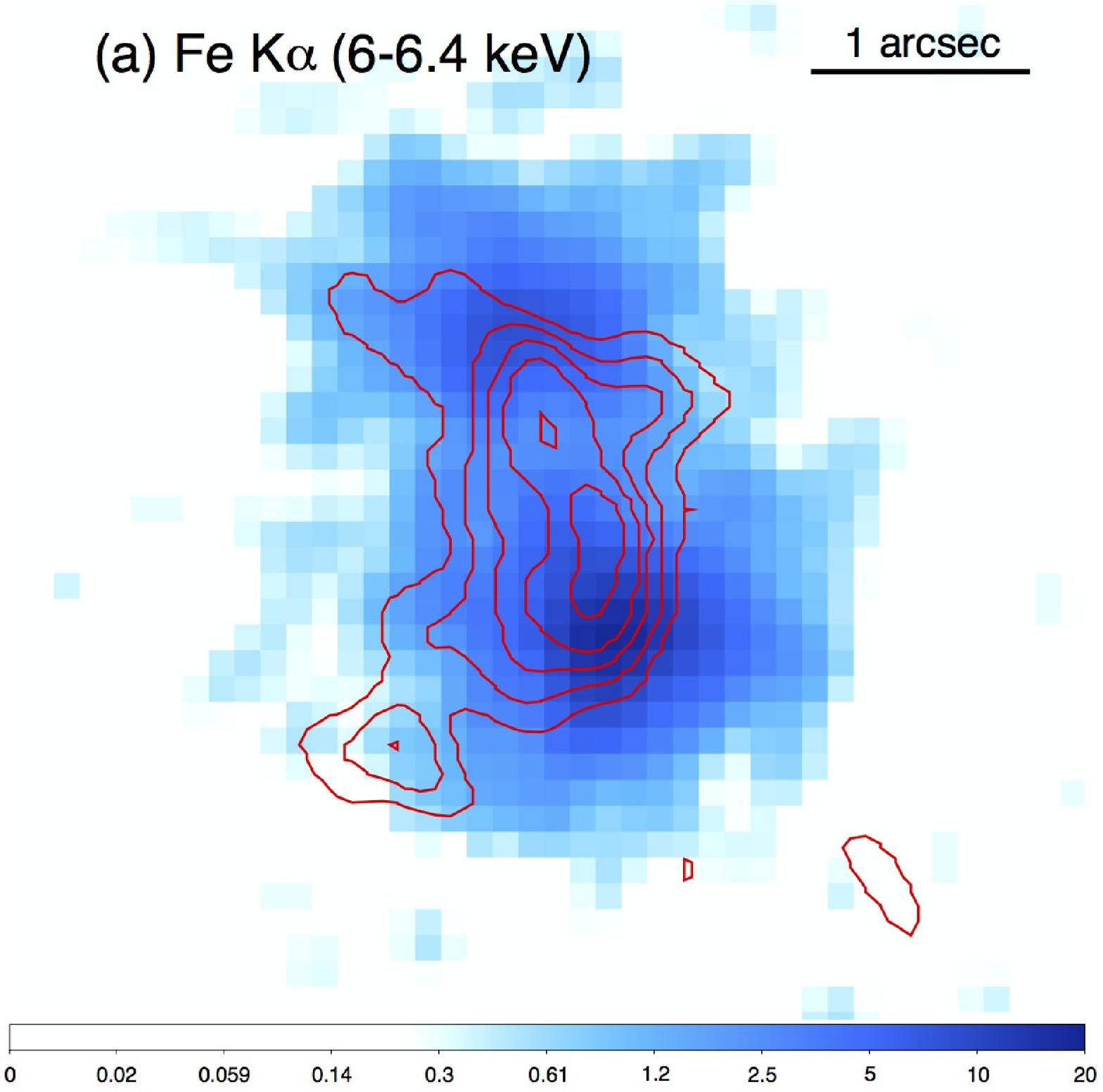}
\plotone{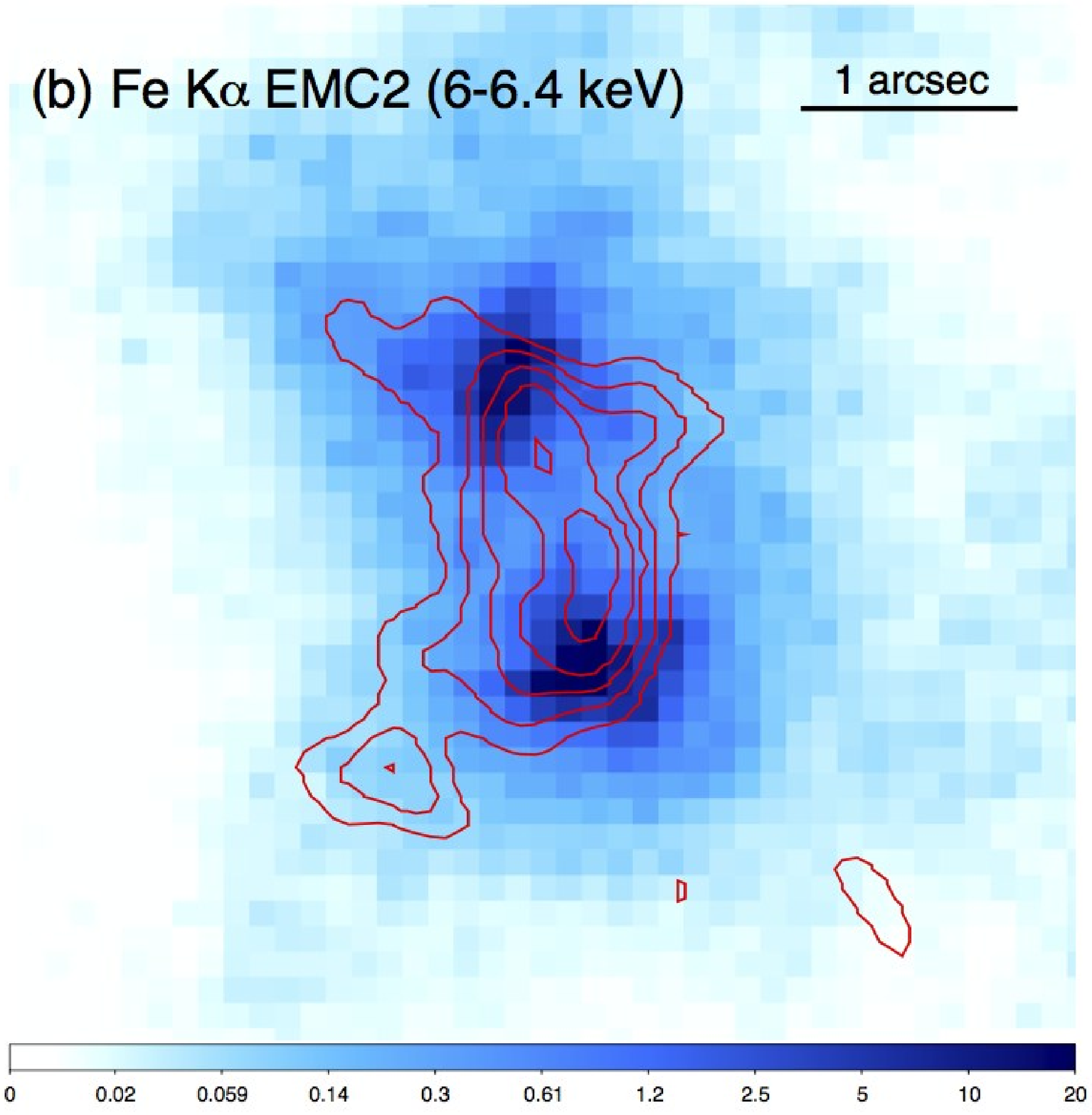}
\epsscale{1}
\caption{(a) The Fe K$\alpha$ line emission image overlaid with the contours
  of SMA CO(3-2) emission (U et al. 2011); (b) PSF-deconvolved
  Fe K$\alpha$ line emission image with the same CO contours.
\label{fig8}}
\end{figure}

\begin{figure}
\centerline{\includegraphics[scale=0.78,angle=-90]{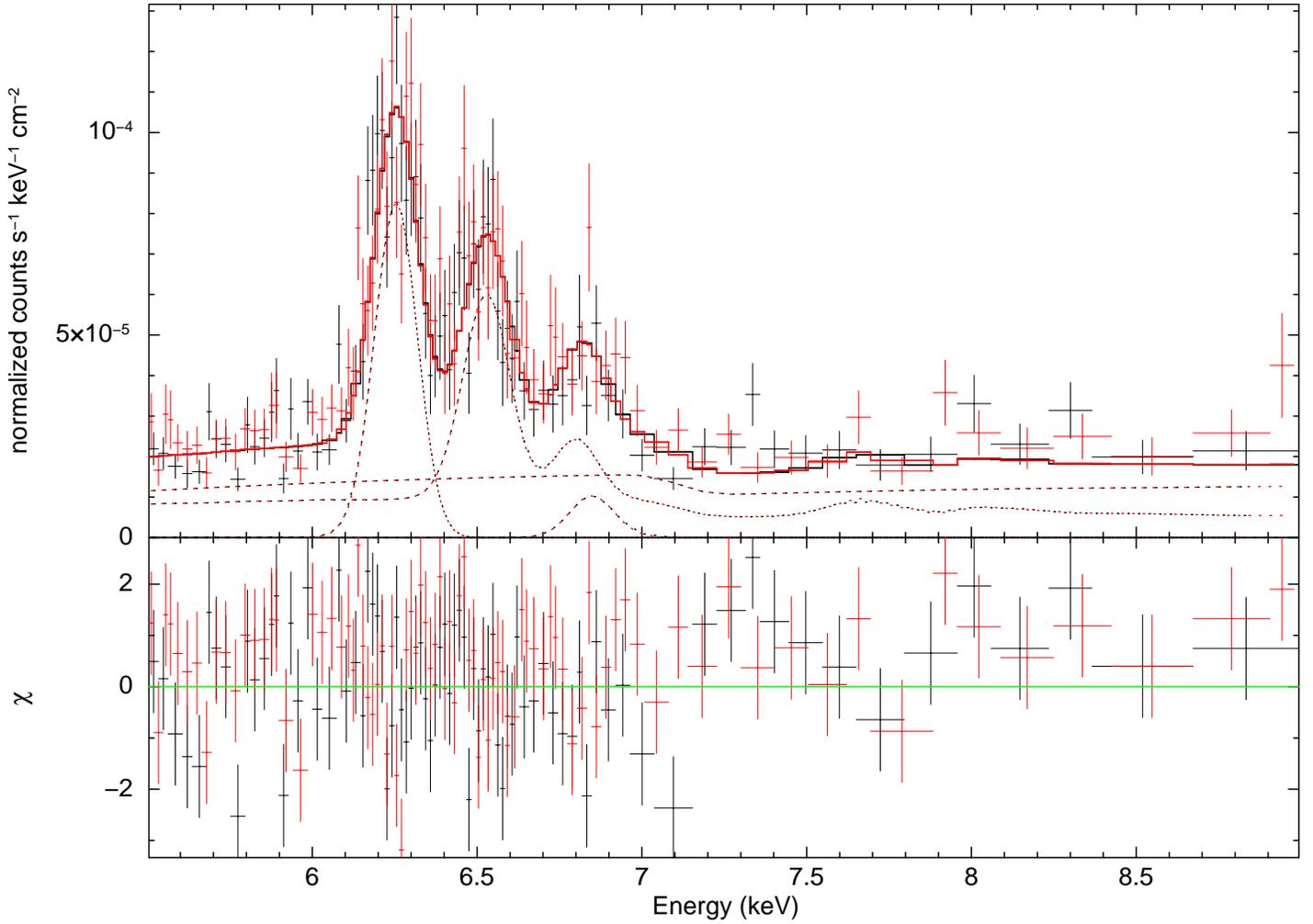}} 
\caption{The X-ray spectra of the NGC 6240 $r=10 \arcsec$ central
  emission (black: ACIS data; red: HETG zeroth order data) and the fit
  using thermal plasma model for the hot gas (see the text
  in \S~\ref{sec:physical}).
\label{fig9}}
\end{figure}

\begin{figure}
\centerline{\includegraphics[scale=0.78,angle=-90]{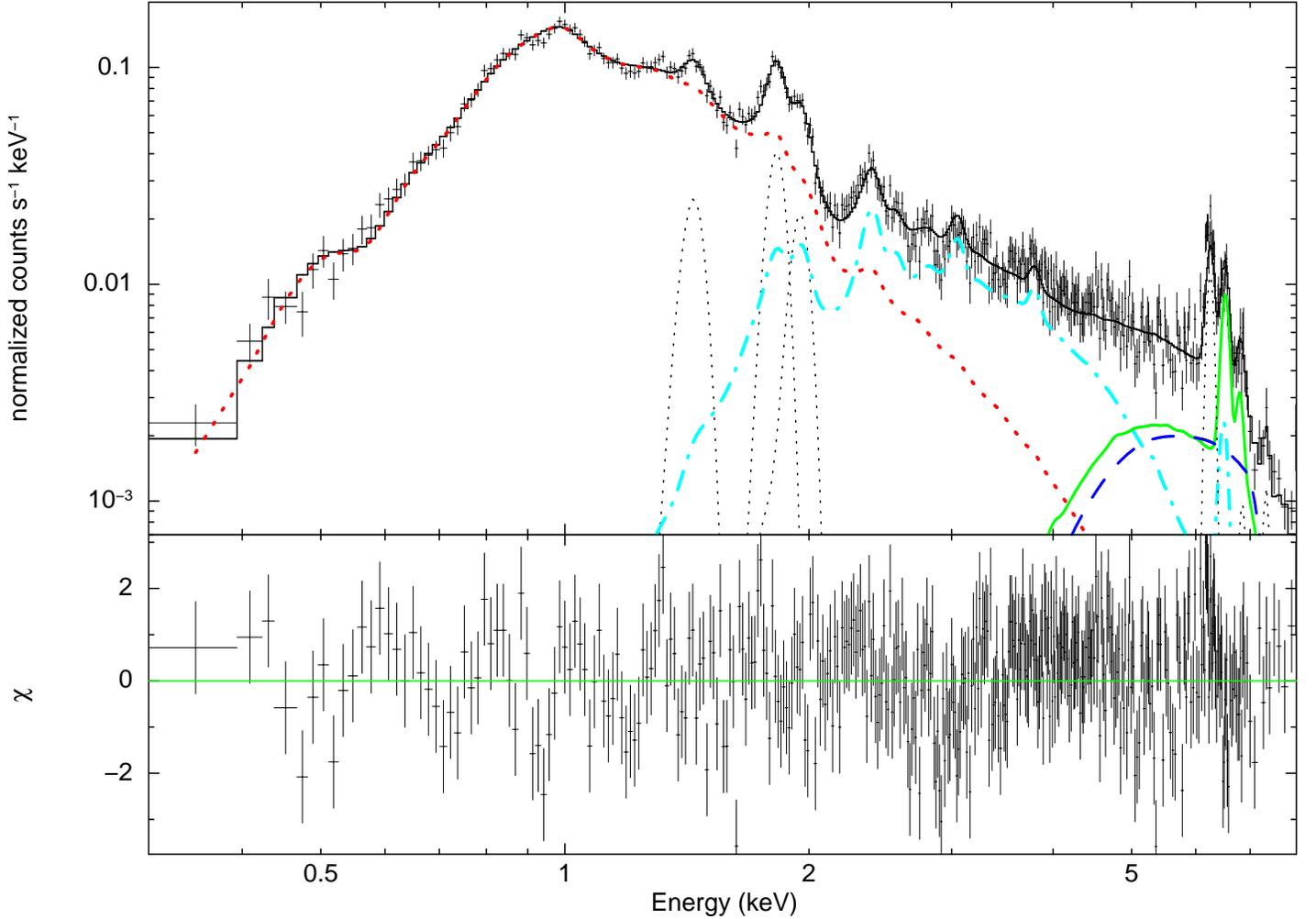}} 
\caption{The broad band (0.3--8 keV) ACIS spectrum of the NGC 6240
  $r=10 \arcsec$ central emission (black line: the total model
  emission). The best fit model components are overplotted including
  three thermal components for the hot gas, two with lower temperature
  (red dotted line: the $kT_1$=1.03 keV component, cyan dash-dotted
  line: $kT_2=1.56$ keV) and one with a high temperature ($kT_3=6.0$
  keV) emitting the \ion{Fe}{25} line (green solid line). The absorbed
  powerlaw component (blue dashed line) represents the any scattered
  nuclear continuum and four narrow gaussian lines correspond to the
  MgXII, SiXIV, SXVI and FeI K$\alpha$ lines (see the text in
  \S~\ref{sec:physical}).
\label{fig9a}}
\end{figure}

However, the morphological comparison showed some interesting
correspondences and differences: (1) in the nuclear region, while
\ion{Fe}{25} and [FeII] both peaks (Figure~\ref{fig6}) at the radio
nuclei, bright H$_2$ emission is observed between the nuclei
(Figure~\ref{fig7}); (2) H$_2$ and the X-ray \ion{Fe}{25} emission
share similar morphology at the `arms' extending southeast and
southwest of the southern nucleus (Figure~\ref{fig7}), which is not
seen with the [FeII] emission.  Why do the near-IR H$_2$ and the X-ray
\ion{Fe}{25} emission share similar extended morphology, but not
    [FeII]?  We suggest that these differences can be attributed to
    two aspects. First, there are presence of shocks in different
    velocity ranges. Secondly, the physical scales of SNR is much
    localized compared to starburst-driven winds.

\citet{vanderWerf93} suggested that the [FeII] emission in NGC 6240 may originate in shocks associated with SNRs during the starburst in the nuclei, based on the
high [FeII]/Br$\gamma$ ratio.  When dust grains are destroyed by fast shocks, the Fe abundance in the gas phase is enhanced.  Thus it is straightforward to expect that the
[FeII] and the highly ionized Fe emission peaks at the nuclei of NGC 6240, where the intense
star formation activities are and young SNR are produced
\citep{Tecza00,Pollack07}.  Note that
      \ion{Fe}{25} in the southern nucleus is $\sim$3 times brighter
      than the northern one (see Figure~\ref{fig7} and
      Table~\ref{tab1}), in agreement with a ratio of $\sim$2 found in
      [FeII] and implying a higher SN rate in the southern nucleus.

From the X-ray perspective, the gas is heated to the high temperature
producing the observed hard X-ray emission by the thermalization of
the kinetic energy from an ensemble of SNe and winds from massive
stars, inside the starburst region, and by internal shocks in the
superwind at larger radii (Tomisaka \& Ikeuchi 1988, Norman \& Ikeuchi
1989, Heckman et al. 1990).  \ion{Fe}{25} emission also peaks at the
the radio nuclei and the peaks of [FeII] emission (Figure~\ref{fig6}).
We can assume that we observe the shocked wind matter, whose
temperature then gives us a measure of the shock velocity. For
non-radiative, strong shocks in a fully ionized monoatomic gas, the
post-shock temperature is $T_{sh}=3\mu v_{sh}^2/16 k$, where $\mu$ is
the mean mass per particle and $k$ is the Boltzmann constant (McKee \&
Hollenbach 1980).  For the observed hot gas ($kT=6.15$ keV, or $\sim
7\times 10^7$~K), we obtain a shock velocity $v_{sh}=2200$ km
s$^{-1}$.  Other than shocks driven by the starburst, such high
velocity shocks are unlikely to be induced by the merger process in
NGC 6240: even in direct collision systems like the Taffy Galaxy (UGC
12914/5), the progenitors collided at $\sim$600 km s$^{-1}$
\citep{Braine03}.  The merging galaxies in NGC 6240 have well passed
the first encounter, without showing morphology like the ring galaxies
exemplifying such a strong direct collision, and the estimated
collision velocity from the CO line width is only between 150 and 300
km s$^{-1}$ \citep{Wang91,Tecza00}.

In contrast, the bright H$_2$ emission observed between the nuclei of
NGC 6240 has been interpreted to originate in global dissipative
shocks resulting from the collisions of the ISM of the merging disk
galaxies \citep{Herbst90,vanderWerf93,Max05}.  The shock velocity was
constrained to be at most 40 km s$^{-1}$ \citep{vanderWerf93}.
However, the fainter large scale `arms' of H$_2$ emission extending SE
and SW of the southern nucleus needs a different explanation, as it
closely follows the X-ray emission.  As fast outflows (several
thousand km s$^{-1}$) driven by the starburst are expected (e.g.,
Heckman et al.\ 1990, Fabbiano et al.\ 1990; see Veilleux et al.\ 2005
for a review), this large scale H$_2$ emission can be interpreted as
originated from molecular material entrained and shocked by the
superwind.  The thermal velocity of this 6 keV gas, estimated as the
adiabatic sound speed $v_{th}=(5kT/3\mu m_{\rm H})^{1/2} \sim$1260 km
s$^{-1}$ (assuming solar abundance), is much higher than the escape
velocity $v_{esc}=\sqrt{2GM/r}$, which is $\sim$210 km s$^{-1}$ at
$r\sim 1$ kpc using the dynamical mass $M_{dyn}=6\times 10^9
M_{\odot}$ in \citet{Tacconi99}.  Therefore the expanding hot bubble
can ``blow out'' to larger radial distance such as the 5 kpc extent
seen here.  Following a simple model of a shock propagating in an
inhomogeneous ISM in \citet{vanderWerf93}, the high velocity shock
($v_1=2200$ km s$^{-1}$, see above) in the low density wind
responsible for the \ion{Fe}{25} emission will slow down significantly
when encountering a layer of cold molecular gas.  The decreased shock
speed $v_2$ can be estimated using the density contrast between the
X-ray gas ($\rho_1 \sim 0.1\eta^{-1/2}$ cm$^{-3}$) and the molecular
cloud ($\rho_2 \sim 10^4-10^5$ cm$^{-3}$), following
$v_2/v_1=\sqrt{\rho_1/\rho_2}$ (Spitzer 1978, Equation [12-39]; note
this is different from the shock jump condition).  The expected
$v_2\sim 7-20$ km s$^{-1}$ agrees well with the velocity constraint
from shock excitation for the observed H$_2$ emission
\citep[e.g.,][]{vanderWerf93}.  We have also shown that the hard X-ray
emission shares similar characteristics with the optical H$\alpha$
emission (Figure~\ref{fig6a}), with a few filaments and loops that are
more apparent in the soft X-rays (Komossa et al. 2003, Wang et
al.\ 2013). The overall picture is that previous models, consisting of
multi-temperature, multi-zone thermal plasma powered by starburst can
explain the observed emission properties given that the outflows (and
consequently shocks) have a large velocity range: the hard X-rays are
associated with fast shocks driven by SNe explosions and the soft
X-rays/H$\alpha$ related to slower shocks associated with mass loaded
starburst winds.

More evidence supporting this scenario also comes from the presence of
high velocity H$_2$ and H$\alpha$ emission in the `arms' region, based
on prominent wings in the H$_2$ 1-0 S(1) line (with full width at zero
intensity of $\sim$1600 km s$^{-1}$; van der Werf et al. 1993; Engel
et al. 2010) and H$\alpha$ line profile \citep{Heckman90}, tracing the
ambient gas entrained and shocked by the superwind.  In a very recent
CO(1-0) interferometry observation of NGC 6240, \citet{Feruglio13}
further identified evidence for a shock wave that propagates eastward
from the nuclei.  One of the resolved structures of blue shifted CO
emission is spatially coincident with the SE H$_2$ filament (Max et
al. 2005), which is also associated with H$\alpha$ and soft X-ray
emission.  The derived kinetic power of the outflowing CO gas is
$6\times 10^{42}$ erg s$^{-1}$, with an estimated age $>2\times 10^7$
years.  This implies a total kinetic energy of $E_{kin}=3.8\times
10^{57}$ ergs.  Recall that the available kinetic input from the
multiple SNe explosions during the starburst is $4\times 10^{58}$ ergs
($10^{51}$ ergs per SN $\times$ 2 yr$^{-1}$ $\times 2\times 10^7$ yr),
there seems no deficit in the energy budget.  However, we cannot yet
draw conclusion that no additional energy injection is required, until
a full tally of the bulk kinetic energy and thermal energy of the
outflowing gas in other phases is available for NGC 6240.  Under the
assumption that the 70 million K gas is volume filling ($\eta=100\%$;
Strickland \& Stevens 2000), its $E_th$ would exceed the available
kinetic energy provided by the recent starburst by $2.5\times 10^{58}$
ergs, implying additional kinetic power input such as from an AGN
outflow or multiple star forming episodes.  In fact, the thermal
energy content in the large scale ($\sim$100 kpc) soft X-ray halo
($kT=0.65$ keV) in NGC 6240 identified by\citet{Nardini13} is
estimated to be $4.9\times 10^{58}$ ergs, which hints at additional
energy input from a widespread, enhanced star formation proceeding at
steady rate over $\sim$200 Myr \citep{Nardini13} besides the most
recent nuclear starburst.

Lastly, we note that a good spatial correlation is also expected if
the H$_2$ emission is excited by localized X-ray irradiation of
molecular clouds \citep{Draine90}.  However, this was concluded
unlikely \citep{vanderWerf93}, since the required SN rate of 7
yr$^{-1}$ \citep{Draine91} is much higher than the inferred rate from
radio, infrared, and our X-ray measurement.
Thus we conclude that propagation of fast shocks originated in the starburst driven wind into the ambient dense gas can account for the large scale X-ray and H$_2$ morphological correspondence.

\section{Conclusions}

Combining new and archival {\em Chandra} observations of NGC 6240, we have
obtained the deepest X-ray image of the central $r=5$ kpc
($10\arcsec$) of the merging ULIRG NGC 6240 with sub-arcsecond
resolution.  Studying the hard X-ray emission centered at Fe K complex
and comparing to images in other wavebands, here for the first time we clearly resolved hard extended emission with its extension established, and assessed the nuclei contribution to the hard emission separately.  Such a hard extended emission was predicted to be the innermost manifestation of superwinds.  So far, the superwind nature of NGC 6240 had been studied mostly in the H${\alpha}$, molecular gas, and soft X-rays.  We summarize our findings as follows:

\begin{enumerate}

\item The X-ray emission from the pair of active nuclei and the
  diffuse hot gas is resolved in detail. Extended X-ray
  emission from $kT\sim 6$ keV ($\sim$70 million K) hot gas was found
  over a spatial scale of 5 kpc, indicating the presence of fast
  shocks with velocity of $\sim$2200 km s$^{-1}$.     

\item The observed luminosity of the 70 million K temperature gas is
  $L_{0.5-8{\rm keV}}=5.3\times 10^{41}$ erg s$^{-1}$, with a total mass of
  $M_{hot}=1.8\times 10^8 M_{\odot}$ and thermal energy
  $E_{th}=6.5\times 10^{57}$ ergs, assuming a filling factor of 1\%
  for the hot gas.  The total iron mass in the highly ionized plasma is 
  $M_{\rm Fe XXV}=4.6\times 10^5 M_{\odot}$.  Both the energetics and
  the iron mass in the hot gas are commensurate with the expected injection from the starburst, assuming an average SN rate of 2 yr$^{-1}$ that is in agreement with its high star formation rate and estimates from non-thermal radio emission and [FeII] luminosity.
  
\item The spatial distribution of the highly ionized gas emitting
  \ion{Fe}{25} peaks together with the [FeII] emission around the radio nuclei,  associated with the high concentration of young SNRs during the nuclear starburst.  The extended \ion{Fe}{25} emission shows a remarkable correspondence to the large scale morphology of H$_2$(1-0) S(1) line emission and H$\alpha$ filaments.  Propagation of fast shocks originated in the starburst driven wind into the ambient dense gas can account for this morphological correspondence.  The Fe I K$\alpha$ emission peaks at the active nuclei. No
  evidence for fluorescent Fe I emission is found in the CO filament
  connecting the two nuclei.

\end{enumerate}

Although the above superwind scenario accounts for the extended hard
X-ray emission and observational evidence in other wavebands, we
cannot yet rule out additional energy injection in the central region
of NGC 6240 (e.g., heating by an AGN outflow).  Given the poorly
constrained volume filling factor of the X-ray emitting gas, the
thermal energy of the hot gas may exceed the available kinetic energy
provided by the SNe explosions during the most recent starburst
($E_{th}-E_{SB}=2.5\times 10^{58}$ ergs for $\eta=100\%$), implying
additional energy input such as from an AGN outflow or more complex
star formation activities besides the recent nuclear starburst (see
discussion in Nardini et al.\ 2013).  A full tally of the bulk kinetic
energy and thermal energy of the outflowing gas in other phases is
crucial for the evaluation of energy budget in the NGC 6240 system. We
will further characterize the soft X-ray emission within the central
15 kpc in this merging galaxy in a forthcoming paper (Wang et al.\ in
preparation).

\acknowledgments

We acknowledge the anonymous referee for the careful reading and
constructive suggestions. This work is supported by NASA grant
GO1-12123X (PI: Fabbiano). We acknowledge support from the CXC, which
is operated by the Smithsonian Astrophysical Observatory (SAO) for and
on behalf of NASA under Contract NAS8-03060. J.W. acknowledges support
from the Northwestern University CIERA Fellowship.  This material is
based upon work supported in part by the National Science Foundation
Grant No. 1066293 and the hospitality of the Aspen Center for Physics.
M.K. is a member of CXC.  Some of the data presented here were
obtained at the W. M. Keck Observatory, which is operated as a
scientific partnership among the California Institute of Technology,
the University of California and NASA. The W. M. Keck Observatory and
the Keck II AO system were made possible by the generous financial
support of the W.M. Keck Foundation.  This research has made use of
data obtained from the {\em Chandra} Data Archive, and software
provided by the CXC in the application packages CIAO. This work has
made use of the Hubble Legacy Archive, which is a collaboration
between the Space Telescope Science Institute (STScI/NASA), the Space
Telescope European Coordinating Facility (ST-ECF/ESA) and the Canadian
Astronomy Data Centre (CADC/NRC/CSA).

{\it Facilities:} \facility{CXO (ACIS)}

\clearpage

\begin{table}[htb]
\begin{center}
\caption{Chandra ACIS Observation Log for NGC 6240.}
\label{tab0}
\footnotesize
\begin{tabular}{cccccc} \hline\hline
ObsID & Date & Exposure (ks) & Instrument &  Data Mode & Roll Angle ($^{\rm o}$)
 \\   \hline
1590 & Feb-22-2002 & 37 & ACIS-S & Faint & 246.2\\
6909 & May-11-2006 & 143 & HETG/ACIS-S & VFaint & 129.9\\
6908 & May-16-2006 & 159 & HETG/ACIS-S & VFaint & 138.1\\
12713 & May-31-2011 & 146 & ACIS-S & Faint & 167.2
\\ \hline
Total Effective Exposure$^{\ast}$ (ks) ... & & 363 & & & 
\\ \hline
\end{tabular}
\end{center}
\begin{flushleft} 
\footnotesize{$^{\ast}$ The effective exposure time for the {\em
    Chandra} HETG zeroth order image refers to the equivalent exposure
  needed if the data were taken with direct {\em Chandra} ACIS
  imaging, calculated by comparing its effective area to that of ACIS
  imaging without grating.}
\end{flushleft}
\end{table}

\clearpage

\begin{table}[htbp]
\begin{center}
\caption{Chandra ACIS Measurements of the Complex Fe K Lines in NGC 6240.}
\label{tab1}
\footnotesize
\begin{tabular}{cccccc} \hline\hline
Region & Counts (5.5-8 keV) &  & Fe~I  & Fe~XXV   & Fe~XXVI
 \\   \hline
       & & Energy (keV) $^{\ast}$   & 6.26$\pm$0.01    &  6.54$\pm$0.03  &  6.86$\pm$0.04  \\  
 Nuc-N & 908 & Flux $^{\ast\ast}$  &  10.7$\pm$0.8 & 2.2$\pm$0.5  &  1.6$\pm$0.6 \\
       & & EW (eV)  &  645$^{+680}_{-18}$  & 73$^{+36}_{-24}$  &  105$^{+134}_{-32}$  \\ 
\hline
       & & Energy (keV) $^{\ast}$   &  6.23$\pm$0.01 & 6.46$\pm$0.01 & 6.68$_{-0.01}^{+0.03}$  \\  
 Nuc-S & 2096 & Flux $^{\ast\ast}$        & 13.5$\pm$0.9  & 8.0$\pm$0.8 & 4.5$\pm$0.7  \\
       & & EW (eV) &  342$\pm$30 & 154$_{-20}^{+15}$  &  115$_{-20}^{+24}$   \\ 
\hline
       & & Energy (keV) $^{\ast}$   & 6.25$\pm$0.01  & 6.51$_{-0.02}^{+0.01}$  & 6.74$\pm$0.02   \\  
 $r\leq 10\arcsec$  & 4856 & Flux $^{\ast\ast}$        &  28.4$_{-1.2}^{+1.3}$ & 16.2$\pm$1.1  &  4.3$_{-1.3}^{+0.8}$  \\
       & & EW (eV) &   359$\pm$22  &  170$_{-4}^{+8}$ &   53$\pm$14    \\ \hline
       & & Energy (keV) $^{\ast}$   & 6.26$\pm$0.02  & 6.52$\pm$0.02  &  6.84$\pm$0.04  \\  
 $r\leq 70\arcsec$ & ... & Flux $^{\ast\ast}$        &  26$\pm$8 &  18$\pm$7  & 9$\pm$7  \\
 ({\em XMM})$^{\dagger}$      & & EW (eV) & 300$\pm$100   & 220$\pm$90  &  120$\pm$90   \\ \hline
\end{tabular}
\end{center}
\begin{flushleft} 
\footnotesize{$^{\ast}$ Observed energies of the line centers.\\
$^{\ast\ast}$ Photon flux in the unit of 10$^{-6}$ photons s$^{-1}$ cm$^{-2}$.\\
$^{\dagger}$ {\em XMM}-Newton EPIC measured FeK line complex from Boller et al.(2003). Because of the lower angular resolution of {\em XMM}-Newton, the spectrum is extracted using a circular region with radius of 70\arcsec.}
\end{flushleft}
\end{table}

\clearpage

\clearpage

\end{document}